\RequirePackage{fix-cm}
\documentclass[twocolumn]{svjour3}          
\smartqed  
\usepackage{graphicx}
\usepackage{amsmath}
\usepackage{amssymb}
\usepackage[linesnumbered,boxed,vlined,algoruled]{algorithm2e} 
\usepackage{textcomp}
\usepackage[misc]{ifsym}
\usepackage[hyphens]{url}
\usepackage{mathptmx}      
\usepackage{cite}
\usepackage{natbib}

\journalname{Soft Computing}

\newcommand{\vertices}{V}
\newcommand{\edges}{E}
\newcommand{\unassigned}{U}

\newcommand{\embeddingalgorithm}{\mathcal{A}}
\newcommand{\graphclass}{\mathcal{G}}

\newcommand{\embedded}{\mathrm{E}_{\text{emb}}}
\newcommand{\leaf}{\mathrm{Leaf}}
\newcommand{\degree}{\mathrm{deg}}

\newcommand{\thres}{\bar{\vertices}}
\newcommand{\tw}{\mathrm{tw}}
\newcommand{\pemb}{p_{\text{emb}}}
\newcommand{\AND}{\,\, \mathrm{and}\,\,}

\renewcommand{\input}{I}
\newcommand{\hardware}{H}
\newcommand{\king}{KG}

\newcommand{\complete}{K}
\newcommand{\graph}{H}
\newcommand{\tree}{\bar{T}}
\newcommand{\ind}{\bar{I}}

\newcommand{\maxiter}{t_{\mathrm{max}}}

\newcommand{\minor}[1]{$(\mathrm{M#1})$}
\newcommand{\treedec}[1]{$\mathrm{(T#1)}$}

\SetKwInOut{Input}{Input}
\SetKwInOut{Output}{Output}
\SetKwInOut{Ensure}{Ensure}
\SetKwInOut{Require}{Require}

\SetKwData{move}{move}
\SetKwData{swap}{swap}
\SetKwData{shift}{shift}
\SetKwData{rf}{random\_float}
\SetKwData{ads}{allow\_any\_direction\_shift}
\SetKwData{schedule}{schedule}
\SetKwData{guidingpattern}{guiding\_pattern}
\SetKwData{true}{true}
\SetKwData{false}{false}
\SetKwData{deletablepatterns}{deletable\_patterns}
\SetKwData{nodelete}{no\_del}
\SetKwData{isdeletable}{is\_deletable}
\SetKwData{bfspath}{bfs\_path}

\begin{document}

\title{Minor-embedding heuristics for large-scale annealing 
processors with sparse hardware graphs of up to 102,400 nodes
\thanks{A preliminary version of this paper has been published in 
\citep{Sug2018}.}
}


\titlerunning{Minor embedding on large-scale annealers}

\author{Yuya Sugie $^{1,2}$ \and
	Yuki Yoshida $^{3}$ [0000-0002-1402-7840] \and
	Normann Mertig $^{1, \text{\Letter}}$ [0000-0003-3025-7141] \and
	Takashi Takemoto $^{1}$ [0000-0002-5949-2252] \and
        Hiroshi Teramoto $^{4,5}$\and
        Atsuyoshi Nakamura $^{2,4}$ [0000-0001-7078-8655] \and
        Ichigaku Takigawa $^{2,4,5,6,7}$ [0000-0001-5633-995X] \and
        Shin-ichi Minato $^{4,8}$\and
        Masanao Yamaoka $^{1}$\and
        Tamiki Komatsuzaki $^{4,6}$ [0000-0001-7175-8474]
}

\authorrunning{Y. Sugie et al.} 

\institute{
	\at $^{1}$ Hitachi Hokkaido University Laboratory, Center for 
Exploratory Research, Research and Development Group, Hitachi, Ltd., Sapporo 
001-0021, Japan\\
	\email{normann.mertig.ee@hitachi.com}
\and
	\at $^{2}$ Graduate School of Information Science and Technology, 
Hokkaido University,	Kita 14 Nishi 9, Kita-ku, Sapporo 060-0814, Japan
\and
	\at $^{3}$ Department of Complexity Science and Engineering, Graduate 
School of Frontier Sciences, The University of Tokyo, 5-1-5 Kashiwanoha, 
Kashiwa, Chiba 	277-8561, Japan
\and
	\at $^4$ Research Center of Mathematics for Social Creativity, Research 
Institute for Electronic Science, Hokkaido University, Kita 20 Nishi 10, 
Kita-Ku, Sapporo 001-0020, Japan
\and
	\at $^5$ PRESTO, Japan Science and Technology Agency (JST), 
Kawaguchi-shi, Saitama 332-0012, Japan
\and
	\at $^6$ Institute for Chemical Reaction Design and Discovery 
(WPI-ICReDD), Hokkaido University, Kita 21, Nishi 10, Kita-ku, Sapporo, 
Hokkaido 001-0021, Japan
\and
	\at $^7$ Center for Advanced Intelligence Project (AIP), RIKEN, 
Nihonbashi 1-chome, Mitsui Building, 15th floor, 1-4-1 Nihonbashi, Chuo-ku, 
Tokyo 103-0027, Japan
\and
	\at $^8$ Graduate School of Informatics, Kyoto University, 
Yoshida-Honmachi, Sakyo-ku, Kyoto 606-8501, Japan
}

\date{Received: date / Accepted: date}

\maketitle
\begin{abstract}
Minor embedding heuristics have become an indispensable tool for compiling 
problems in quadratically unconstrained binary optimization (QUBO) into the 
hardware graphs of quantum and CMOS annealing processors.
While recent embedding heuristics have been developed for annealers of moderate 
size (about 2000 nodes) the size of the latest CMOS annealing processor (with 
102,400 nodes) poses entirely new demands on the embedding heuristic.
This raises the question, if recent embedding heuristics can maintain meaningful 
embedding performance on hardware graphs of increasing size.
Here, we develop an improved version of the probabilistic-swap-shift-annealing 
(PSSA) embedding heuristic [which has recently been demonstrated to outperform 
the standard embedding heuristic by D-Wave Systems \citep{CaiMacRoy2014}] and 
evaluate its embedding performance on hardware graphs of increasing size.
For random-cubic and Bar{\'a}basi-Albert graphs we find the embedding 
performance of improved PSSA to consistently exceed the threshold of the best 
known complete graph embedding by a factor of $3.2$ and $2.8$, respectively, up 
to hardware graphs with 102,400 nodes.
On the other hand, for random graphs with constant edge density not even 
improved PSSA can overcome the deterministic threshold guaranteed by the 
existence of the best known complete graph embedding.
Finally, we prove a new upper bound on the maximal embeddable size of complete 
graphs into hardware graphs of CMOS annealers and show that the embedding 
performance of its currently best known complete graph embedding has optimal 
order for hardware graphs with fixed coordination number.

\end{abstract}

\section{Introduction}
\label{sec:Introduction}

The last decade has witnessed impressive progress in the development of a new 
hardware architecture, which is commonly known as annealing processor.
This development was sparked by the discovery of a new model for quantum 
computation \citep{KadNis1998,Far2001}, which led to the development of 
annealing hardware based on superconducting quantum bits \citep{Joh2011, DWave}.
Simultaneously, the performance of hardware processors executing simulated 
annealing \citep{KirGelVec1983} has improved considerably, e.g., due to 
algorithmic advances \citep{Isa2015, ZhuOchKat2015, Ara2018} as well as 
customized hardware circuits based on CMOS or laser technology, 
see \citep{Yam2016, Oku2017, Fuj2017, Fuj2018, AC2018, Tak2019} or 
\citep{Ina2016, McMah2016} for references.
While hardware realizations may differ in their implementation, all annealing 
processors perform exactly the same task.
That is, they provide a fast method for finding the ground state configuration, 
which minimizes the energy of an Ising model.
In that, they represent the ideal hardware for solving combinatorial 
optimization problems in quadratic unconstrained binary optimization (QUBO) 
\citep[see, e.g.,][]{Luc2014, Koc2014} and related applications, e.g., in 
quantum chemistry \citep{XiaBiaKai2017} or machine learning \citep{Nev2009}.

\begin{figure*}[tb]
	\centering{\includegraphics[width=0.85\textwidth]{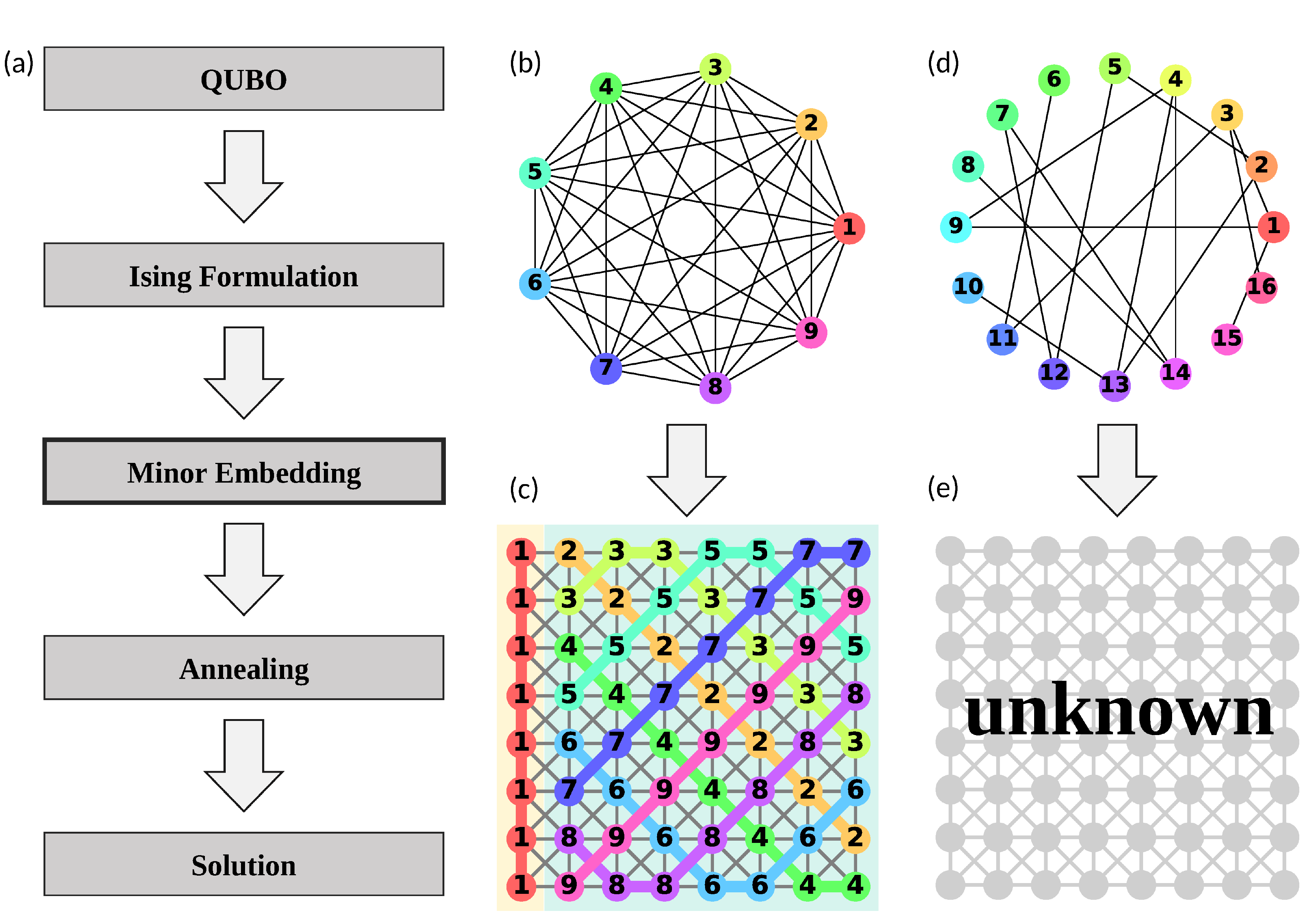}}
	\caption{(a) Work flow for solving QUBO problems on annealing hardware, 
emphasizing the role of graph-minor embeddings for representing (b, d) the 
spin-spin connectivity graphs of a given QUBO input on (c, e) annealing hardware 
with sparse spin-spin connectivity. (b, c) Minor embedding of a complete graph 
$\complete_{9}$ into a King's graph $\king_{8,8}$. Each super vertex is marked 
by the corresponding vertex label of the input graph.}
	\label{fig:workflow}
\end{figure*}

A general problem in quadratically unconstrained binary optimization results in 
Ising models which consist of $N$ binary spin variables $\left \{\sigma_{i} 
\right\}$ which take values $\sigma_{i}\in\{-1,1\}$, $i=1,\cdots,N$.
The energy of each spin configuration is given by
\begin{equation}
\label{eq:IsingModel}
H(\sigma_{i}) = -\sum_{i,j=1}^{N} \sigma_{i} J_{ij} \sigma_{j} -  
\sum_{i=1}^{N} 
h_{i} \sigma_{i},
\end{equation}
where $h_{i}\in\mathbb{R}$ and $J_{ij}\in\mathbb{R}$ are externally fixed model 
parameters, known as magnetic fields and spin-spin couplings, respectively.
These model parameters encode the energy landscape (or cost function) of the 
optimization problem.
Unfortunately, present day quantum annealers can only provide for a finite 
amount of couplings between spins \citep{Joh2011, DWave}, resulting in tangible 
restrictions on $J_{ij}$.
In particular, the non-zero $J_{ij}$ of a quantum annealer induce the famous 
hardware topology known as chimera graph.
Similarly, large-scale CMOS annealing processors with up to 102,400 spins and 
fast parallel updates have only been realized with sparse hardware topologies 
\citep{Yam2016, Tak2019, AC2018}.
In this case, the non-zero spin-spin couplings $J_{ij}$ induce the hardware 
topology of a King's graph, as illustrated in Fig.~\ref{fig:workflow}.
In order to solve QUBO problems on such annealing processors, the ground state 
search of the given Ising model has to be mapped to a given hardware by means of 
minor embedding \citep{Cho2008}.
See Fig.~\ref{fig:workflow}(a) for a general work flow.
In minor embedding each spin of the original Hamiltonian is represented by a 
tightly connected group of spins on the hardware, which are forced to point 
into the same direction, see Fig.~\ref{fig:workflow}(c).
Such a group of hardware spins is called a \textit{super vertex}, representing 
a single spin of the original input problem.
In contrast to a single spin, a super vertex can be adjacent to many other super 
vertices on the hardware.
Precisely this property allows for encoding the spin-spin couplings of the 
original Ising model between super vertices, see Fig.~\ref{fig:workflow}(b, c) 
for an example.

Minor embedding for annealing processors, i.e., embedding an input graph 
$\input$ (induced by the non-zero couplings of a given optimization problem) 
into a hardware graph $\hardware$ (induced by the non-zero couplings of the 
hardware) has attracted significant attention over the past decade.
In particular, for optimization problems whose input graph $\input$ posses a 
predefined structure fast deterministic and often optimal embeddings into the 
static hardware graphs $\hardware$ are known.
See \citep{Cho2011, Ven2014, KlySulHum2014, Boo2016, Oku2016} for embeddings of 
complete graphs, \citep{Oku2016, Goo2018} for embeddings of bipartite graphs, or 
\citep{Zar2017} for embeddings of Cartesian product graphs.
On the other hand, many problems in combinatorial optimization, such as solving 
the clique problem on social networks, produce sparse input graphs whose 
connections appear random.
Such optimization problems do not posses a tangible structure which can 
systematically be exploited for minor embedding.
Yet, embedding such problems by resorting to minor embeddings of complete 
graphs seems wasteful.
For example, for the presently best quantum annealer with 2048 hardware spins, 
it would result in representing input problems with only 65 spins.
Similarly, for the currently largest CMOS annealer with 102400 spins, it would 
result in representing input problems with only 321 spins.
Hence, improving over the embeddability threshold ensured by the best known 
complete graph embeddings is desirable.

To this end, one could, in principle, resort to the exact algorithms invented 
by \citep{Rob1995} and later improved by \citep{Kaw2009} and \citep{Adl2010}.
However, these algorithms are not constructive and have prohibitively large 
runtime
\begin{equation}
 \mathcal{O}\left(2^{(2k + 1)\log{k}} \left|\vertices(\input)\right|^{2k} 
2^{2\left|\vertices(\input)\right|^2} \left|\edges(\hardware)\right|\right),
\end{equation}
which scales exponentially with the number of vertices of the input graph 
$|\vertices(\input)|$, linearly in the number of edges on the hardware graph 
$|\edges(\hardware)|$, and further depends on the branch width $k$ of the 
hardware graph.
For this reason, it is important to have efficient heuristics which find graph 
minors with high probability, rather than attempting an exhaustive search or 
proving minor exclusion.
Such a heuristic was, for example, proposed in \citep{CaiMacRoy2014} and has 
subsequently become state of the art due to its inclusion in the standard 
software package provided by D-Wave Systems.
An alternative heuristic, called probabilistic-swap-shift-annealing (PSSA), was 
subsequently invented by YY as a submission to the ``Hokkaido University \& 
Hitachi 2nd New-Concept Computing Contest 
2017.''\footnote{\url{https://hokudai-hitachi2017-2.contest.atcoder.jp/}}
There it showed outstanding performance as compared to hundreds of other 
submissions, eventually winning the contest.
Its superior performance to \citep{CaiMacRoy2014} was later demonstrated by the 
authors in \citep{Sug2018}.
While both heuristics have initially been developed for annealers of moderate 
size (about 2000 spins) the release of the latest CMOS annealing processor with 
102,400 spins poses entirely new demands on the performance of the embedding 
heuristics.
With D-Wave Systems doubling the number of qubits of its annealers roughly every 
two years a similar demand on embedding heuristics is expected for quantum 
annealers in the near future.

In this paper, we present an improved version of PSSA and use it in a case study 
to evaluate the performance of embedding heuristics on ever larger hardware 
graphs of up to 102,400 spins.
The algorithmic improvements of PSSA include (i) an additional search phase, 
(ii) a degree-oriented super-vertex shift rule, and (iii) optimized annealing 
schedules.
The performance of improved PSSA is then investigated with respect to various 
types of random input graphs on hardware graphs of increasing size.
An embedding performance consistently exceeding the embedding threshold of the 
best known complete graph embedding by a constant factor $c$ is observed for 
random-cubic ($c=3.2$) and Bar{\'a}basi-Albert ($c=2.8$) graphs for hardware 
graphs up to 102,400 spins.
Incidentally this constitutes an average performance gain of 42\% and 28\%, 
respectively, over the previously published version of PSSA \citep{Sug2018}.
On the other hand, for sparse random graphs with constant edge density even the 
embedding performance of improved PSSA shrinks to the deterministic threshold 
guaranteed by the existence of the best known complete graph embedding on large 
hardware graphs.
Finally, this paper provides a new upper bound on the maximal complete graphs 
embeddable into a King's graph hardware of present CMOS annealing processors.
In addition, we prove that the baseline performance of the currently best known 
complete graph embedding has optimal order for hardware graphs with fixed 
coordination number.

This paper is organized as follows.
In Sect.~\ref{sec:Preliminaries} we lay out the basic notation, define graph 
minors, and specify the algorithm task.
In Sect.~\ref{sec:PSSA}, we briefly outline PSSA as previously published in 
\citep{Sug2018} in order to make the paper self-contained.
In Sect.~\ref{sec:Improvements} we present several improvements of PSSA, 
including (i) an additional terminal search phase, (ii) a degree-oriented super 
vertex shift rule, and (iii) optimized annealing schedules.
Finally, in Sect.~\ref{sec:Scalability} we evaluate the embedding performance 
of the improved PSSA on hardware graphs of increasing size (up to 102,400 
nodes) with respect to various input problems.
In Sect.~\ref{sec:CliqueMinorScaling} we prove an upper bound on the maximal 
complete graphs embeddable into a hardware King's graphs of CMOS annealers and 
show that the baseline performance of the currently best known complete graph 
embedding has optimal order for hardware graphs with fixed coordination number.
We summarize in Sect.~\ref{sec:Summary} and discuss open questions for future 
research.

\section{Preliminaries}
\label{sec:Preliminaries}

In this section we introduce some basic notations and definitions.
In particular, we define minor embeddings and super vertex placements, describe 
the hardware graph, specify the algorithm task, and discuss several input graphs 
which will be used for evaluating the embedding performance of improved PSSA on 
hardware graphs of ever larger size.

\subsection{Graph minor and super vertex placement}
\label{sec:Graphminor}

In what follows we consider undirected and simple graphs.
Let $\input=(\vertices(\input), \edges(\input))$ denote an input graph.
Its vertex set $\vertices(\input)$ represents the spin indices $i=1,\cdots,N$ of 
the original QUBO problem, Eq.~\eqref{eq:IsingModel}.
Its edges 
$(i,j)\in\edges(\input)\subset\vertices(\input)\times\vertices(\input)$ are 
induced by the non-zero entries of the corresponding symmetric connectivity 
matrix $J_{ij}$, $i,j=1,\cdots,N$.
Let $\hardware=(\vertices(\hardware), \edges(\hardware))$ denote a hardware 
graph.
Its vertex set $\vertices(\hardware)$ shall represent the spins of the 
hardware, while its edges $\edges(\hardware)$ are induced by the non-zero 
couplings of the annealing processor.

We now define a minor embedding, similar to \citep{CaiMacRoy2014}.
Our definition will proceed in two stages.
In the first stage we define a super vertex placement.
\begin{definition}[Super Vertex Placement]
  Let $\input$, $\hardware$ be two graphs. A super vertex placement is a 
function $\phi:\vertices(\input)\rightarrow 2^{\vertices(\hardware)}$ which 
assigns each vertex $i\in\vertices(\input)$ to a subset of vertices 
$\phi(i)\subset\vertices(\hardware)$, such that:
\begin{itemize}
  \item[] \hspace{-0.5cm}\minor{1} $\forall i \in \vertices(\input)$: 
     $\phi(i)\neq\emptyset$ and the subgraph induced by $\phi(i)$ in 
$\hardware$ is connected.
  \item[] \hspace{-0.5cm}\minor{2} $\forall i,j\in \vertices(\input)$ with 
$i\neq j$: $\phi(i)\cap\phi(j)=\emptyset$.
\end{itemize}
\end{definition}
In the second stage we define a minor embedding.
\begin{definition}[Minor Embedding]
Let $\input$, $\hardware$ be two graphs. A super vertex placement $\phi$ is a 
minor embedding, if in 
addition to property \minor{1} and \minor{2} we have
\begin{itemize}
  \item[] \hspace{-0.5cm}\minor{3} $\forall (i,j)\in\edges(\input)$:\\ $\exists 
(u,v)\in\phi(i)\times\phi(j)$ such that $(u,v)\in\edges(\hardware)$. 
\end{itemize} 
\end{definition}
In what follows we refer to each set $\phi(i)$, induced by a super vertex 
placement  $\phi$, as a super vertex.
In order for $\phi(i)$ to represent a single spin from the input problem, we 
require super vertices to be non-empty and connected, hence \minor{1}.
In addition, a single spin of the hardware cannot represent multiple spins of 
an input problem, hence \minor{2}.
Finally, a valid representation of a given Ising model can be provided, if and 
only if all edges of the input graph $\input$, induced by non-zero values of 
$J_{ij}$, can be represented by at least one edge between the corresponding 
super vertices on the hardware graph.
Hence, a given super vertex placement $\phi$ is a suitable hardware mapping, if 
and only if it is a minor embedding.
See Fig.~\ref{fig:workflow}(b, c) for an example.

In order to rate the quality of a super vertex placement $\phi$, we define a 
function which counts the edges that a super vertex placement faithfully 
represents on the hardware graph.
\begin{definition}[Number of Embedded Edges]
Let $\input$, $\hardware$ be two graphs and $\phi$ a super vertex placement. 
The number of edges embedded by $\phi$ is defined as
\begin{align}
\label{eq:embedded}
\embedded(\phi) :=& \left|\left\{(i,j)\in\edges(\input)\,\,|\,\,
                    \exists u\in\phi(i), v\in\phi(j) \right.\right.\\
                    \nonumber
              \phantom{=}& \hspace{2.4cm}\left.\left. \text{with } (u,v) \in 
                                              \edges(\hardware)\right\}\right|.
\end{align}
\end{definition}
For a general super vertex placement we have that $\embedded(\phi) \le 
|\edges(\input)|$ and a valid minor embedding, satisfying condition \minor{3}, 
is found, if and only if $\embedded(\phi)=|\edges(\input)|$.

Note that for certain inputs $\input$ and a fixed hardware $\hardware$, it may 
be the case that no minor embedding exists.
In that case the number of embedded edges will be strictly smaller than the 
number of edges of the input graph, such that $\embedded(\phi) < 
|\edges(\input)| $ for any super vertex placement $\phi$.
In this case the heuristic we propose in the next sections is bound to fail.
Furthermore, throughout the annealing phase of PSSA, we restrict our 
implementation to super vertex placements where each super vertex $\phi(i)$ can 
be parametrized by a path.
We denote the endpoints of the vertex path, i.e., the set of its leaves by 
$\leaf{[\phi(i)]}$.

Finally, we emphasize that finding a minor embedding is the key ingredient for 
mapping QUBO problems, Eq.~\eqref{eq:IsingModel}, onto the spin-spin couplings 
and external magnetic fields of an annealing processor.
Prior to finding the minor embedding one determines the input graphs $\input$ 
from the non-zero entries of the matrix $J_{i,j}$.
After a minor embedding is found, the the spin-spin couplings and external 
magnetic fields of the annealing processor can be determined from the input 
parameters and the minor embedding ($J_{i,j}$, $h_{i}, \phi$) as described in 
\citep{Cho2008}.\footnote{Roughly speaking, (i) the coupling $J_{i,j}$ will be 
applied to exactly one coupler connecting the super vertices $\phi(i)$ and 
$\phi(j)$, (ii) The internal connections of super vertex $\phi(i)$ are set to a 
strength of order $\mathcal{O}(\sum_{j\in nbr(i)}|J_{i,j}|-|h_{i}|)$, where 
$nbr(i)$ denotes the vertex neighbors of $i\in\vertices(\input)$, and (iii) the 
magnatic field $h_{i}$ is distributed across the spins of the super vertex 
$\phi(i)$.}
Both processes are straight forward and will no further be mentioned in this 
paper.

\subsection{Hardware graphs}
\label{sec:HardwareGraphs}

The development of annealing processors with sparse hardware topology currently 
knows two pertinent hardware architectures.
One is the Chimera graph topology adopted by quantum annealers \citep{Cho2011, 
KlySulHum2014}.
The other is the King's graph topology adopted by the CMOS annealing processors 
\citep{Oku2016, Oku2017}.
Both hardware graphs have very similar general features, i.e.:\
(i) both graphs have a fixed coordination number,
(ii) for both graphs the tree-width grows as a square root of vertices 
$\mathcal{O}(\sqrt{|\vertices(\hardware)|})$, and
(iii) for both hardware graphs constructive minor embeddings of complete 
graphs ensure the embeddability of input graphs with 
$\mathcal{O}(\sqrt{|\vertices(\hardware)|})$ vertices.
(See next section for details.)

In this study, we will focus on hardware topologies $\hardware$ forming a 
square King's graph \citep{Oku2017} while omitting Chimera-type hardware graphs 
entirely.
We made this choice due to the following arguments:\
(i) The large scale hardware structures on which we benchmark the performance of 
improved PSSA has so far only been realized for CMOS annealers with hardware 
King's graphs of up to 102,400 spins.
(ii) A performance analysis of PSSA on Chimera graphs has previously been 
published in \citep{Sug2018} and is no longer the main objective of this paper.
(iii) We believe that our evaluation of the embedding performance of PSSA on 
hardware King's graphs of increasing size would not fundamentally change, if 
carried out on Chimera graphs due to the similarity of both graphs with respect 
to general features.

A square King's graph represents all valid moves of the king chess piece on a 
chessboard, i.e., each vertex represents a square of the chessboard and each 
edge is a valid move.
We denote the King's graph by the symbol $\king_{L,L}$, where $L$ denotes the 
width of the chessboard and $L\times L$ denotes the total number of vertices 
$|\vertices(\hardware)|$.
See Fig.~\ref{fig:workflow}(c, e) for an example.
In what follows, the King's graph $\king_{L,L}$ defined by the hardware is 
usually fixed.
The goal is to find minor embeddings for large input graphs which have as many 
vertices $\left|\vertices(\input)\right|$ as possible.

\subsection{Best known complete graph embedding}
\label{sec:CliqueMinors}

A simple and fast baseline strategy for minor embedding into a fixed hardware 
graph exploits the existence of known complete graph embeddings.
In particular, if a minor embedding of a complete graph $\complete_{N}$ with 
$N$ vertices is known, embedding of other input graphs with $N$ vertices is 
trivial.
For a fixed King's graph hardware $\king_{L,L}$, embedding complete graphs 
$\complete_{N}$ with up to $N=L+1$ vertices, is always possible using the 
construction of \citep{Oku2016}.
See Fig.~\ref{fig:workflow}(b, c) for a sketch.
Hence, minor embedding of any graph $\input$ with $|\vertices(\input)|\le L+1$ 
is trivial.
On the other hand, a systematic embedding of complete graphs $\complete_{N}$ 
with $N>L+1$ is currently unknown and believed to be impossible.
(A proof showing that complete graphs $\complete_{N}$ with $N>2L$ are not 
embeddable into King's graph $\king_{L,L}$ is given in 
Sect.~\ref{sec:CliqueMinorScaling}.)
In absence of a proof of optimality we refer to the embedding of complete graphs 
according to \citep{Oku2016} as the \textit{best known complete graph 
embedding}.

\subsection{Sparse random input graphs}
\label{sec:InputGraph}

While the best known complete graph embedding does not allow for embedding 
complete graphs with $|\vertices(\input)|> L+1$ into a King's graph, going 
beyond this embedding threshold should be feasible, if the input graph is 
sparse.
See Fig.~\ref{fig:workflow}(d, e).
PSSA tries to find minor embeddings precisely for this kind of input, in order 
to widen the range of QUBO problems amenable to the hardware of annealing 
processors.
However, if a certain QUBO problem induces sparse input graphs $\input$ with a 
predefined structure, it is advisable to resort to a deterministic minor 
embedding.
This is usually faster and most likely produces embeddings for larger input 
graphs than a general purpose heuristic.
This strategy has, for example, been applied for the minor embedding of 
bipartite graphs \citep{Goo2018, Oku2016} or the minor embedding of Cartesian 
product graphs \citep{Zar2017}.
In that, an embedding heuristic should be applied to sparse input problems whose 
structure is not known in advance.
In other words, an embedding heuristics is used on input graphs which appear 
to be random to some degree.
Such randomized sparse input graphs can appear, e.g., in graph coloring or when 
solving the clique problem \citep{Luc2014} on social network graphs 
\citep{Alb2002}.
To benchmark PSSA on a wide variety of potential input types, this paper 
considers the following three famous classes of random input graphs.
(i) Random cubic graphs as a model of super low edge density,
(ii) Bar{\'a}basi-Albert random graphs \citep{Alb2002} as a prototype of 
scale-free graph structures from social network science, and
(iii) Erd\H{o}s-R{\'e}nyi random graphs \citep{Erd1959, Bol1985} with constant 
edge density as a prototype of a sparse random graph with high complexity.
For a detailed description of these graphs, the reader is referred to 
Appendix~\ref{app:experimentalconditions}.

\subsection{Embedding probability and embedding threshold}

We now define our performance measures.

\begin{definition}[Embedding Probability]
\label{eq:embeddingprobability}
Let $\graphclass$ denote a class of input graphs $\input\in\graphclass$, e.g., 
random cubic graphs.
Let $\embeddingalgorithm$ denote an embedding algorithm such as PSSA.
For a fixed hardware graph $\hardware$ we define the embedding probability
\begin{equation}
 \pemb(|\vertices(\input)|, \hardware, \graphclass, \embeddingalgorithm) 
\nonumber
\end{equation}
as the ratio of input samples from graph class $\graphclass$ restricted to 
graphs of vertex size $|\vertices(\input)|$ for which the embedding algorithm 
$\embeddingalgorithm$ finds a minor embedding.
\end{definition}
Note that the probabilistic nature of the embedding probability originates both 
from the potentially probabilistic elements of the embedding algorithm as well 
as the probabilistic nature of the input graphs.
Further note that, the definition of the embedding probability may in principal 
be augmented by including further dependencies, such as the number of edges 
$|\edges(\input)|$ of an input graph.
In this paper, we chose the number of vertices $|\vertices(\input)|$ because it 
directly corresponds to the number of spins of the input problem.
In addition, for all graph classes considered in this paper the number of edges 
$|\edges(\input)|$ immediately follows from $|\vertices(\input)|$.

In this paper we will evaluate the embedding probability on a fixed hardware 
graph $\hardware$ for ever larger sizes $|\vertices(\input)|$ of the input 
graphs $\input \in \graphclass$.
Up to fluctuations this typically results in a monotonically decreasing function 
with high embedding probability at small sizes of $|\vertices(\input)|$ and 
zero probability for large sizes of $|\vertices(\input)|$.
See Fig.~\ref{fig:bfs} for a preview.
We then define the embedding threshold as follows.
\begin{definition}[Embedding Threshold]
Let $\hardware$ be a fixed hardware graph.
Let $\graphclass$ denote a class of input graphs $\input\in\graphclass$.
Let $\embeddingalgorithm$ be an embedding algorithm and 
$\pemb(|\vertices(\input)|, \hardware, \graphclass, \embeddingalgorithm)$ the 
corresponding embedding probability.
For a fixed constant $p$ with $0<p\le1$, we define the embedding threshold
\begin{align}
 \thres(\hardware, \graphclass, \embeddingalgorithm, p) =& \\ \nonumber
  \phantom{=}& \hspace{-0.5cm} \min \{|\vertices(\input)|\in\mathbb{N} \,\, | 
\,\, \pemb(|\vertices(\input)|, \hardware, \graphclass,                          
                          \embeddingalgorithm) < p\},
\end{align}
as the minimal vertex size $|\vertices(\input)|$ for which the embedding 
probability $\pemb$ falls below a prescribed threshold $p$.
\end{definition}
Throughout this paper we will often denote the embedding probability as 
$\pemb(|\vertices(\input)|)$ and the embedding threshold as $\thres(L)$ 
because:\
(i) The graph class $\graphclass$ (random cubic, Bar{\'a}basi-Albert, or 
Erd\H{o}s-R{\'e}nyi) will always be specified from the context.
(ii) The embedding algorithm $\embeddingalgorithm$ will always be some version 
of PSSA whose details are specified from the context.
(iii) The hardware graph $\hardware$ will always be a King's graph 
$\king_{L,L}$, whose size is either specified from context or through the side 
length $L$.
The latter is convenient since the embedding threshold is often found to be a 
linear function of the side length $L$ empirically.
(iv) We fix the value of $p$ at high and constant probability 
$p=0.95$.
(The precise value of $p$ is usually not important, since the transition of 
$\pemb(|\vertices(\input)|)$ from 1 to 0 as a function of increasing 
$|\vertices(\input)|$ is usually quite sharp.)

\subsection{Performance target and evaluation methodology}

From the preceding discussion it is obvious that a good heuristic should never 
fall below the embedding threshold ensured by the best known complete graph 
embedding.
Furthermore, a good heuristic should be able to maintain an embedding threshold 
which exceeds the best known complete graph embedding as the size of the 
hardware graph increases.
In particular, on a King's graph we want a heuristic which can maintain an 
embedding threshold $\thres(L)>L+1$ as $L$ increases.

In this paper we will demonstrate that improved PSSA can maintain an embedding 
threshold $\thres(L)>L+1$ even on large hardware graphs for certain types of 
input graphs, such as random cubic and Bar{\'a}basi-Albert graphs.
On the other hand, we also show that not even improved PSSA can beat the 
embedding threshold of the best known complete graph embedding for 
Erd\H{o}s-R{\'e}nyi graphs as the size of the hardware increases.
To corroborate these statements we will now introduce PSSA as previously 
described in \citep{Sug2018} in the next section.
Subsequently, we introduce improved PSSA and tune it on hardware King's graphs 
with $320\times320$ spins, as described in Sect.~\ref{sec:Improvements}.
Finally, we evaluate the embedding threshold of improved PSSA phenomenologically 
on King's graphs of ever larger size in Sect.~\ref{sec:Scalability}.

\section{Probabilistic-Swap-Shift-Annealing (PSSA)}
\label{sec:PSSA}

In order to make this paper self-contained, we outline the core elements of the
probabilistic-swap-shift-annealing heuristic, previously published in 
\citep{Sug2018}.

%
\begin{algorithm}[tb]
  \Input{Input graph $\input$ and hardware graph $\hardware$}
  \Output{Super vertex placement $\phi_{best}$}
  \Ensure{$|\vertices(\input)| \leq |\vertices(\hardware)|$, $|\edges(\input)| 
\leq |\edges(\hardware)|$}
  \Require{Function $\embedded(\phi)$, Eq.~\eqref{eq:embedded}, $\maxiter$, 
\schedule, \guidingpattern}

  \BlankLine
  \tcp{prepare initial placement of super vertices}
  $\phi \gets $ \guidingpattern divided into $|\vertices(\input)|$ super 
vertices\tcp*[l]{see Fig.~\ref{fig:pssa}(a)} $\phi_{best} \gets \phi$; 
\textbf{if} $\embedded(\phi_{best})=|\edges(\input)|$ \Return $\phi_{best}$ and 
terminate\tcp*{minor found}

  \BlankLine
  \tcp{improve super vertex placement through simulated annealing}
  \For{$t=0$ to $\maxiter$}{
    \move $\gets$  \swap or \shift, randomly selected according to \schedule \;
    \uIf(\tcp*[h]{see Fig.~\ref{fig:pssa}(b)}){\move is \swap}{
      $ i,k \gets (i,k)\in \edges(\input)$, randomly selected\;
      $ j \gets j \in \vertices(\input)$ with $\phi(j)$ neighboring $\phi(k)$ 
in 
$\hardware$, randomly selected\;
      $\phi_{proposed} \gets \phi$ with $\phi(i)$ and $\phi(j)$ swapped\;
    }
    \ElseIf(\tcp*[h]{see Fig.~\ref{fig:pssa}(c, d)}){\move is \shift}{
      $i, u \gets i \in \vertices(\input)$ with $|\phi(i)| > 1$ and $u \in 
\leaf[\phi(i)]$, both randomly selected\;
      $\ads \gets$ \true or \false according to \schedule \;
      \uIf(\tcp*[h]{see Fig.~\ref{fig:pssa}(d)}){\ads}{
	$j, v \gets j\in\vertices(\input)$, $v\in\leaf[\phi(j)]$ with $v$ 
adjacent to $u$ in $\hardware$, randomly selected\;
      }\Else(\tcp*[h]{see Fig.~\ref{fig:pssa}(c)}){
	$j, v \gets j\in\vertices(\input)$, $v\in\leaf[\phi(j)]$ with $v$ 
adjacent to $u$ along \guidingpattern, randomly selected\;
      }
      $\phi_{proposed} \gets \phi$ with $u$ deleted from $\phi(i)$ and assigned 
to $\phi(j)$\;
    }

    \tcp{evaluate acceptance of proposed move}
    $\mathrm{\Delta E} \gets$ $\embedded(\phi_{proposed}) - \embedded(\phi)$; 
$T(t) \gets$ temperature according to \schedule \;
    \If(\tcp*[h]{accept and update}){$\exp (\mathrm{\Delta E} / T(t)) > $ \rf $ 
\in \left[ 0, 1 \right)$}{
      $\phi \gets \phi_{proposed}$\;
      \If{$\embedded(\phi_{best}) < \embedded(\phi)$}{
	$\phi_{best} \gets \phi$\;
	\textbf{if} $\embedded(\phi_{best})=|\edges(\input)|$ \Return 
$\phi_{best}$ and terminate\tcp*{minor found}
      }
    }
  }
  
  \BlankLine
  \Return $\phi_{best}$ \tcp{even if $\embedded(\phi_{best}) < 
|\edges(\input)|$, i.e., minor not found}

  \caption{PSSA \citep{Sug2018}}
  \label{alg:pssa}
\end{algorithm}
\begin{figure*}[tb]
	\includegraphics[width=\textwidth]{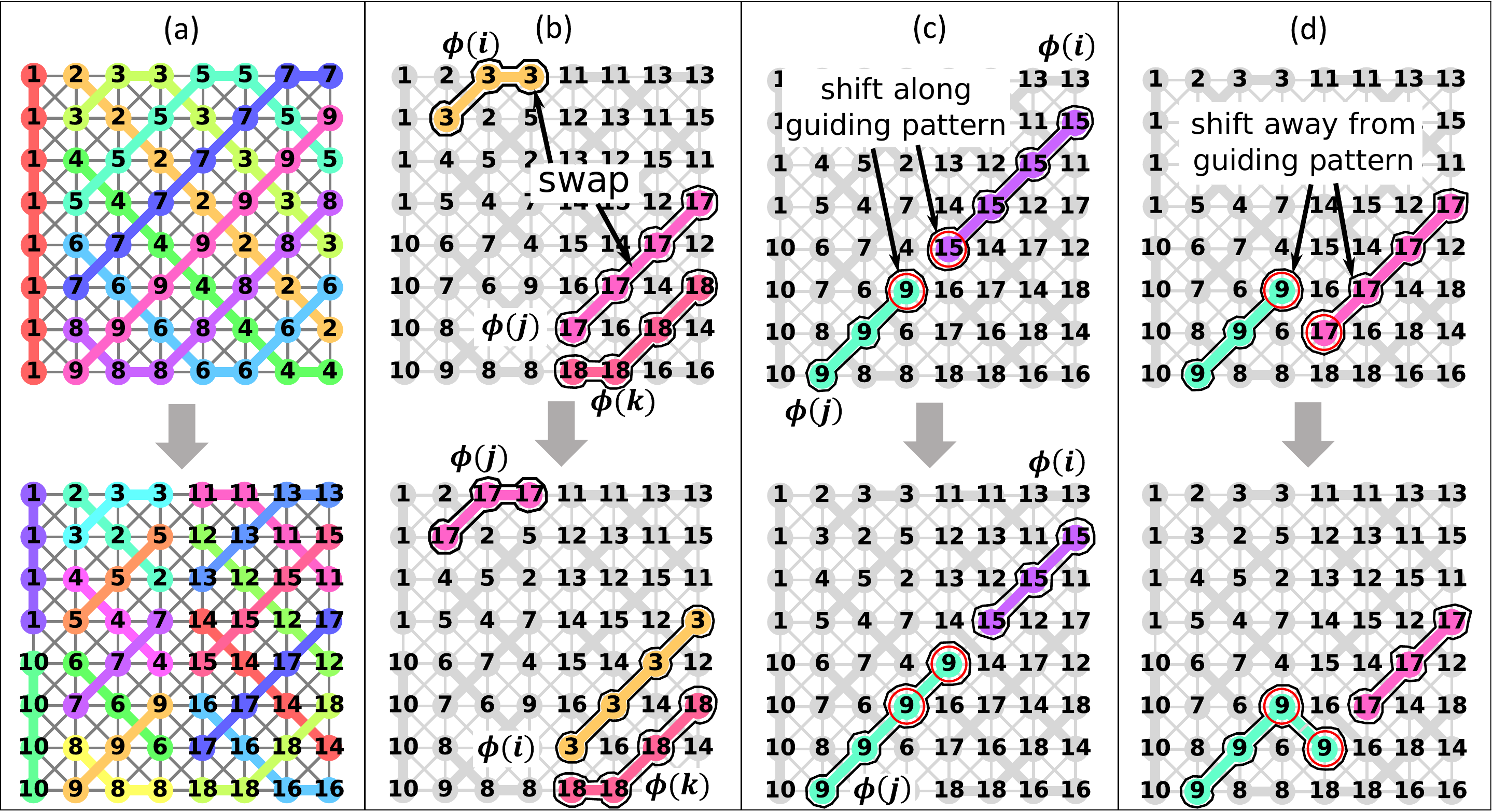}
	\caption{Visualizing the main components of PSSA on a King's graph 
$\king_{8,8}$. (a) The guiding pattern induced by the best known complete graph 
embedding of $\complete_{9}$ and the division of its super vertices for initial 
placement. (b) Swapping of super vertices. (c, d) Shifting the leaves of a 
super 
vertex to its neighbor (c) along and (d) away from the guiding pattern. Each 
super vertex is marked by the corresponding vertex label of the input graph.}
	\label{fig:pssa}
\end{figure*}
For a given input graph $\input$ and a given hardware $\hardware$ PSSA tries to 
find a minor embedding by implementing the following general framework:\
(i) First, an \textit{initial super vertex placement} representing the vertices 
of the input graph $\input$ on the hardware $\hardware$ is prepared.
See Fig.~\ref{fig:pssa}(a).
In general, the initial super vertex placement will not embed all edges of the 
input graph such that $\embedded(\phi) < |\edges(\input)|$.
In this case PSSA proceeds to the annealing search phase.
(ii) In the annealing phase PSSA will successively propose new super vertex 
placements.
To this end it either \textit{swaps} two super vertices, see 
Fig.~\ref{fig:pssa}(b), or \textit{shifts} hardware vertices from one super 
vertex to its neighbor, see Fig.~\ref{fig:pssa}(c, d).
The proposed super vertex placement is accepted, if the number of embedded 
edges $\embedded(\phi)$ grows.
On the other hand, a proposed super vertex placement is accepted with finite 
probability, even if the number of embedded edges decreases.
This avoids trapping the algorithm on incomplete super vertex placements which 
maximize $\embedded(\phi)$ locally.
(iii) The annealing search phase terminates, if a super vertex placement 
representing all edges of the input graph, $\embedded(\phi) = |\edges(\input)|$, 
i.e., a valid minor embedding is found.
Alternatively, PSSA may terminate unsuccessfully after reaching a maximum 
amount of prescribed iterations $\maxiter$, because a minor embedding was 
either not found or does not exist.

In addition to the general framework, PSSA is currently implemented with the 
following specifications:\
(i) PSSA uses super vertices $\phi(i)$ which are parametrized as path on the 
hardware.
(ii) \textit{Initial super vertex placements} are generated by splitting the 
super vertices of the best known complete graph embedding on $\hardware$ into 
$|\vertices(\input)|$ super vertices of almost equal size to represent the 
vertices of the input graph $\vertices(\input)$ on the hardware $\hardware$.
See Fig.~\ref{fig:pssa}(a) for an example on a King's graph.
Incidentally, this guarantees that PSSA never falls below the embedding 
threshold of the best known complete graph embedding.
In addition, this initial placement generates a super vertex placements where 
each super vertex has many neighbors resulting in a high connectivity of super 
vertices.
This is expected to facilitate finding a minor embedding in the successive 
annealing search phase.
(iii) A \textit{swap} is implemented by randomly selecting an edge $(i,k) \in 
\edges(\input)$ and swapping the super vertex $\phi(i)$ with a super vertex 
$\phi(j)$, adjacent to $\phi(k)$ on $\hardware$.
See Fig.~\ref{fig:pssa}(b).
(iv) A shift is implemented by randomly selecting a super vertex $\phi(i)$ 
(with $\left|\phi(i)\right|>1$) and one of its leaf nodes $u\in\leaf[\phi(i)]$.
The shift proposal is completed by deleting $u$ from $\phi(i)$ and attaching it 
to a leaf $v$ of a neighboring super vertices $\phi(j)$ on $\hardware$.
See Fig.~\ref{fig:pssa}(c, d).
If a neighboring leaf $v$ does not exist, the shift proposal is skipped and the 
algorithm proceeds to the next proposal.
If there are multiple candidates for $v$ the algorithm randomly selects one of 
the available nodes with equal probability.
(v) PSSA further uses the super vertices of the best known complete graph 
embedding as a guiding pattern in order to distinguish two types of shift moves.
See top of Fig.~\ref{fig:pssa}(a) for an example of the guiding pattern.
A shift move is along the guiding pattern, if the leaf $u\in\leaf[\phi(i)]$ is 
attached to a leaf $v\in\leaf[\phi(j)]$ with both $u$ and $v$ belonging to the 
same super vertex of the best known complete graph embedding.
See Fig.~\ref{fig:pssa}(c).
On the other hand, a shift move is away from the guiding pattern, if the leaf $u 
\in\leaf[\phi(i)]$ is attached to a leaf $v\in\leaf[\phi(j)]$ with $u$ and $v$ 
belonging to distinct super vertices of the best known complete graph embedding.
See Fig.~\ref{fig:pssa}(d).
PSSA favors shifts along the guiding pattern, in order to prioritize super 
vertices with diagonal orientation.
This is expected to increase the connectivity of the super vertices, and thus, 
to prevent PSSA from getting trapped in a local maximum of the score function 
$\embedded(\phi)<|\edges(\input)|$.
For more details the reader is referred to the pseudo-code summary of improved 
PSSA, given in Algorithm~\ref{alg:pssa}.

\textit{Schedule} -- The original implementation of PSSA divides the annealing 
time into two search phases of equal length.
During both phases the temperature is initialized at a finite value and then 
decreased to zero
\begin{eqnarray}
T(t) &= \begin{cases}
T_{0} \times \left(1 - \frac{2t}{\maxiter}\right)	  & \quad \text{if } 0 
\le t < \frac{\maxiter}{2},\\
T_{\frac{\maxiter}{2}} \times \left(2 - \frac{2t}{\maxiter}\right) & \quad 
\text{if } \frac{\maxiter}{2} \le t \le \maxiter,
\end{cases}
\end{eqnarray}
allowing for a finite acceptance of suboptimal moves in the beginning of each 
phase, while suppressing them towards the end of each phase.
In the first search phase conventional PSSA suppresses shift moves which lead 
away from the guiding pattern.
This policy is expected to protect PSSA from being trapped in suboptimal super 
vertex placements.
If the first search phase fails to produce a valid minor embedding, PSSA enters 
the second phase.
During that second phase we consider a wider search space by allowing for a 
higher proportion of shifts away from the guiding pattern.
To this end, PSSA schedules shifts with probability $p_{s}(t)$ and swaps with 
probability $1-p_{s}(t)$.
If a shift is proposed, an arbitrary shift direction is allowed with probability 
$p_{a}(t)$, while shifts along the guiding pattern are guaranteed with 
probability $1-p_{a}(t)$.
Both $p_{s}(t)$ and $p_{a}(t)$ probabilities are scheduled with simple linear 
schedules.
The detailed scheduling parameters used in numerical experiments are summarized 
in the Appendix~\ref{app:experimentalconditions}.
For further comments on scheduling the reader is referred to 
Sect.~\ref{sec:ImprovedAnnealingSchedules} on improved annealing schedules.

\section{Improvements of PSSA}
\label{sec:Improvements}

In order to improve the performance of PSSA on large input graphs, we 
implemented several modifications which shall be described in the following.
These include (i) an additional terminal search phase, (ii) a degree-oriented 
super-vertex shift rule, and (iii) optimized annealing schedules.

\subsection{Terminal search phase}

If standard PSSA fails to produce a valid minor embedding after $\maxiter$ 
iterations, it usually returns a super vertex placements $\phi_{best}$ 
(henceforth referred to as $\phi$ for brevity) with two pertinent properties:\
(i) $\phi$ occupies all vertices $u \in \vertices(\hardware)$ of the hardware.
This can potentially lead to unnecessary assignments $u \in \phi(i)$ which 
are neither needed for preserving the connected structure of a super vertex 
$\phi(i)$ nor for connecting super vertices $\phi(i)$, $\phi(j)$ with $(i,j)\in 
\edges(\input)$.
Even worse, these unnecessary super vertex assignments may obstruct potential 
connections between super vertices $\phi(i)$, $\phi(j)$ with $(i,j)\in 
\edges(\input)$.
(ii) The super vertex placement $\phi$ returned by PSSA usually represents most 
edges of the input graph already ($\embedded(\phi)\lesssim|\edges(\input)|$).
Thus, finding a few more path connecting super vertices $\phi(i)$, $\phi(j)$ 
with $(i,j)\in \edges(\input)$ may already result in a valid minor embedding 
which fulfills \minor{3}.
Hence, the terminal search tries to transform $\phi$ into a valid minor 
embedding by addressing properties (i) and (ii) as follows.
For details see the pseudo-code summary in Algorithm~\ref{alg:bfs}.
%
\begin{algorithm}[tb]
  \Input{Input graph $\input$ and hardware graph $\hardware$}
  \Output{Super vertex placement $\phi$}
  \Require{PSSA, $\isdeletable(\cdot)$, $\bfspath(\cdot)$}
  \BlankLine
  \tcp{get super vertex placement from PSSA}
  $\phi \gets$ PSSA($\input, \hardware$)\;
  \BlankLine
  \tcp{--- Start terminal search ---}
  \BlankLine
  \tcp{Create free hardware vertices}
  \BlankLine
  Init: $\unassigned \gets \emptyset$ \tcp{Set of free vertices}
  Init: $u \gets 0$ \tcp{hw vertex $u \in \vertices(\hardware) = \{0, 
\cdots, |\vertices(\hardware)|-1\}$}
  Init: $\nodelete \gets 0 $ \tcp{Vert. scanned since last del.}
  \BlankLine
  \While{$\nodelete < (|\vertices(\hardware)|-1)$}{
    \tcp{scan hardware vertices $u = \{0, \cdots, |\vertices(\hardware)|-1\}$}
    $i\gets \phi^{-1}(u)$ \tcp{get preimage of $u$}
    \If{
      $(u\notin\unassigned)\AND(\isdeletable(u,i))$
    }{
      \tcp{shift $u$ to set of free vertices}
      $\phi(i) \gets \phi(i)\setminus\{u\}$\;
      $\unassigned \gets \unassigned \cup \{u\}$\;
      $\nodelete \gets 0$\;
    }\Else{
      $\nodelete \gets \nodelete + 1$\;
    }
    $u \gets (u+1) \mod |\vertices(\hardware)|$ \tcp{check next hw vertex}
  }

  \BlankLine
  \tcp{Find new super vertex links by breadth first search on graph 
induced by free vertices}
  \BlankLine
  \For{$i=0,\cdots,|\vertices(\input)|-1$}{
    \If{
      $\exists (i,j)\in\edges(\input)$ with $\phi(i), \phi(j) \subset 
\hardware$ not connected 
    }{
       \tcp{search BFS path from $\phi(i)$ to $\phi(j)$}
       $\phi, \unassigned \gets \bfspath(i,j)$
    }
  }

  \BlankLine
  \Return $\phi$ \tcp{even if $\embedded(\phi) < |\edges(\input)|$, i.e., minor 
not found}
  \caption{Improved PSSA with terminal search}
  \label{alg:bfs}
\end{algorithm}

\textit{Creating free hardware vertices} -- 
First the hardware vertices $u \in \vertices(\hardware) = \{0, \cdots, 
|\vertices(\hardware)|-1\}$ are iterated over repeatedly and it is checked by 
the subroutine $\isdeletable(\cdot)$, if $u$ can be removed from its 
corresponding super vertex $\phi(i) \ni u$.
In this step a vertex $u$ is removable from $\phi(i)$, if
(p1) removing $u$ from $\phi(i)$ does not destroy the non-empty connected 
structure of the super vertex $\phi(i)$ \minor{1} and
(p2) removing $u$ from $\phi(i)$ does not decrease the number of embedded edges 
$\embedded(\phi)$.
Vertices $u \in \vertices(\hardware)$, which have been removed from a super 
vertex are collected in a set of \textit{free} vertices
\begin{equation}
\label{eq:unassigned}
 \unassigned = \vertices(\hardware) \setminus 
\left(\bigcup_{i\in\vertices(\input)}\phi(i)\right).
\end{equation}
The cleanup process is terminated, if no more vertices $u \in 
\vertices(\hardware)$ can be removed from their super vertex $\phi(i)\ni u$ 
after a sweep of the whole hardware.
For a possible implementation of the subroutine $\isdeletable(\cdot)$ the 
reader is referred to Appendix~\ref{app:Implementation}.

\textit{Super vertex links from breadth first search (BFS)} --
After the cleanup phase the free hardware vertices $u\in\unassigned$ are used 
to create representations of edges $(i,j)\in\edges(\input)$ whose super 
vertices $\phi(i), \phi(j)$ are not yet linked on the hardware.
As described in Algorithm~\ref{alg:bfs} this is done by successively iterating 
through the vertices $i=0,\cdots,|\vertices(\input)|-1$ of the input graph and 
checking for edges $(i,j)\in\edges(\input)$ whose super vertices $\phi(i), 
\phi(j)$ are not yet linked on the hardware.
If such an edge is found the terminal search makes a call to the subroutine 
$\bfspath(\cdot)$ which tries to link up the super vertices $\phi(i), \phi(j)$.
In this step breadth first search \citep{Cor2009} is used on the graph induced 
by the free vertices $\unassigned$ on the hardware graph $\hardware$ to search 
for a path connecting the pair of super vertices  $\phi(i)$, $\phi(j)$.
If such a path is found, the corresponding vertices are included in $\phi(i)$ 
and deleted from $\unassigned$.
The algorithm then proceeds to the next vertex pair $\phi(i)$, $\phi(j)$.
For a possible implementation of the subroutines $\bfspath(\cdot)$ the reader 
is referred to Appendix~\ref{app:Implementation}.

\begin{figure}[tb]
  \centering{\includegraphics[width=\linewidth]{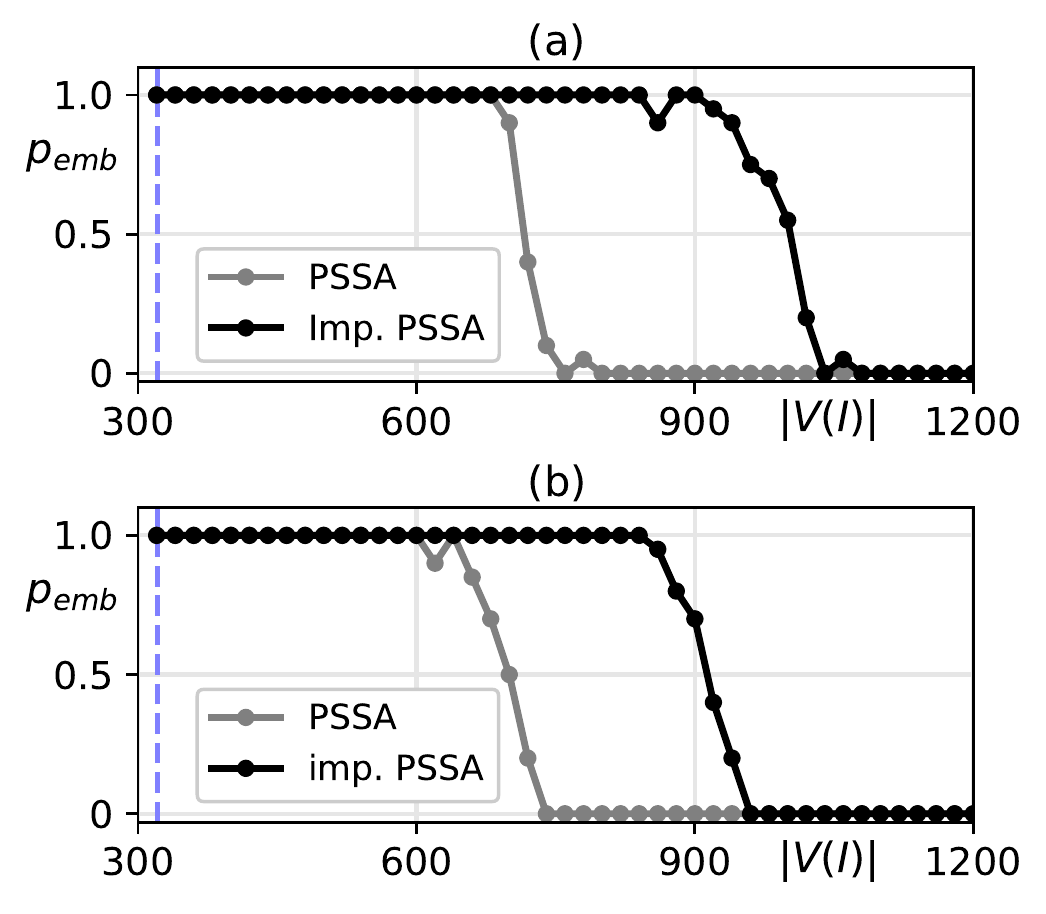}}
  \caption{Comparing the empirical embedding probability (20 inputs) of PSSA 
(gray) and improved PSSA with terminal search (black) for (a) random cubic and 
(b) Bar{\'a}basi-Albert type input graphs on a King's graph $\king_{320,320}$ 
hardware. A (dashed) vertical line indicates the embedding threshold $321$ of 
the best known complete graph embedding.}
	\label{fig:bfs}
\end{figure}

Note that the terminal search algorithm has already been a part of the original 
PSSA as designed by YY.
However, since the terminal search phase had almost no impact on the embedding 
probability on King's graphs of moderate size ($L=52$) its details have 
previously been omitted in \citep{Sug2018}.
On the other hand, for large hardware King's graphs ($L=320$) the terminal 
search phase does have a profound impact on the embedding performance.
This is depicted in Fig.~\ref{fig:bfs} which compares the embedding performance 
of standard and the improved PSSA, showing that the improved PSSA allows for 
increasing the size of the embeddable input problems both for random cubic and 
Bar{\'a}basi-Albert type input graphs $\input$.
For Erd\H{o}s-R{\'e}nyi-type input graphs with 20\% edge density even the 
improved PSSA fails to construct minor embeddings for inputs beyond the 
embeddability threshold ($|\vertices(\input)|>321$) ensured by the existence of 
the best known complete graph embedding.
Reasons for the difficulty of embedding Erd\H{o}s-R{\'e}nyi graphs will be 
discussed in Sect.~\ref{sec:Summary}.

We close this section with a short discussion:\
(i) Note that the terminal search phase will never decrease the number of 
embedded edges $\embedded(\phi)$.
(ii) While PSSA produces super vertex placements $\phi$ which occupy the whole 
hardware, the terminal search phase may leave hardware vertices unused.
Similarly, while standard PSSA produces super vertices $\phi(i)$ which induce a 
path in $\hardware$, this property may be destroyed in the terminal search 
phase due to the deletion of vertices $u\in\vertices(\hardware)$ as well as 
growing additional path.
(iii) Breadth first search is an efficient algorithm for finding the shortest 
path between two vertices and two super vertices can be connected by including 
all the vertices of the path into one super vertex.
Searching for the shortest path is beneficial since the number of nodes needed 
to 
connect two super vertices is smallest.
However, there is no guarantee that the shortest path between two super 
vertices is the optimal choice.
In particular, there may be cases, where the shortest path connecting a pair of 
super vertices may obstruct the paths connecting another pair of super 
vertices, which could be avoided by choosing a longer path.
For this reason the total number of links produced by the terminal search phase 
may depend on the order in which links between super vertex pairs are created 
by breadth first search.
In a similar manner, the deletion of vertices $u$ from their super vertex 
$\phi(i)$ and thus the set $\unassigned$ of free vertices may depend on the 
order in which the deletion of $u$ is tried.
Ultimately, it is not even guaranteed that the super vertex placement $\phi$ 
produced by PSSA is necessarily the input which produces the largest amount of 
embedded edges in the terminal search phase.

\subsection{Degree-weighted shift proposals}

\begin{figure}[tb]
  \centering{\includegraphics[width=\linewidth]{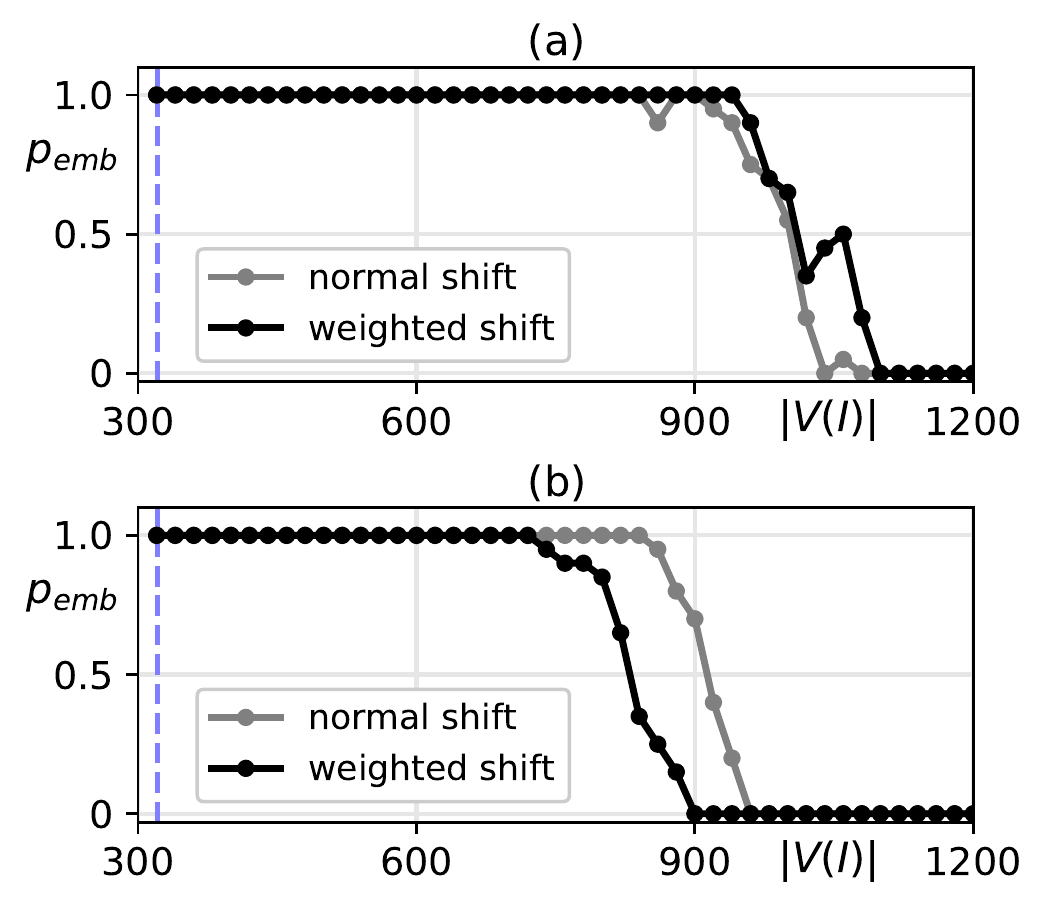}}
  \caption{Comparing the empirical embedding probability (20 inputs) of 
improved PSSA with (black) and without (gray) degree-weighted shift proposals 
for (a) random cubic and (b) Bar{\'a}basi-Albert type input graphs on a King's 
graph $\king_{320,320}$ hardware. A (dashed) vertical line indicates the 
embedding threshold $321$ of the best known complete graph embedding.}
	\label{fig:shift}
\end{figure}

We further improve the PSSA embedding performance, by exploiting the degree 
distribution of the original input graph $\input$.
The basic idea is that vertices with a large amount of neighboring vertices 
should correspond to large super vertices comprising many nodes of the 
hardware, while nodes with few neighboring vertices should correspond to tiny 
super vertices, comprising only few vertices on the hardware.
Most of our attempts to include the information on the degree distribution, in 
particular when creating the initial super vertex placement, went unsuccessful 
and shall not further be reported.
However, including the information of the degree distribution to bias shift 
proposals, gave mild improvements on the embedding probability of random cubic 
graphs, see Fig.~\ref{fig:shift}, and shall be described in more detail.

To begin with, the degree $\degree(i)$ of a vertex $i\in\vertices(\input)$ is 
defined as the number of edges incident to the vertex $i$.
On the other hand, the size of the corresponding super vertex $\phi(i) \subset 
\vertices(\hardware)$ shall be denoted as $|\phi(i)|$.
The degree-weighted shift rule is then applied as follows.
First, we schedule two neighboring leaf nodes $u,v$ with $u \in \phi(i), v \in 
\phi(j)$ for a shift move exactly as in conventional PSSA.
Subsequently, we compute the degree ratio's, which we define as
\begin{align}
 dr(x) :=& \frac{\left|\phi(x)\right|}{\degree(x)} \quad 
x=i,j\in\vertices(\input),
\end{align}
as a measure to which extent the vertex size matches the degree of the input 
graph.
Finally, we propose assigning $u\in\phi(i)$ to $\phi(j)$ with probability
\begin{align}
 \mu(u\in\phi(i)\to u\in\phi(j)) := \frac{dr(i)}{dr(i) + dr(j)},
\end{align}
or alternatively propose shifting $v\in\phi(j)$ from $\phi(j)$ to $\phi(i)$.
This biases the shift proposal to assign a leaf node to the super vertex with 
lower degree ratio.

In order to evaluate the impact of the degree-weighted shift proposal, we 
compared the embedding performance of improved PSSA using the conventional 
shift rule and the degree-weighted shift rule.
In both cases we applied the terminal search phase.
The embedding performance for a King's graph hardware $\king_{320,320}$ is shown 
in Fig.~\ref{fig:shift}, for (a) random cubic and (b) Bar{\'a}basi-Albert-type 
input graphs.
The degree-weighted shift rule shows mild improvements for random cubic graphs.
Surprisingly, it clearly decreases the embedding probability for 
Bar{\'a}basi-Albert graphs, for which its design was originally intended.
Finally, it has no effect on the embedding probability of Erd\H{o}s-R{\'e}nyi 
graphs with 20\% edge density.
Those remain embeddable only by resorting to the best known complete graph 
embedding of $\complete_{321}$.

\subsection{Improved annealing schedules}
\label{sec:ImprovedAnnealingSchedules}

As a last step towards designing an improved PSSA, capable of embedding even 
more edges of a given type of random input graphs, we tried to optimize the 
functional form of the annealing schedules.
In particular, we tested four different functional forms of the temperature 
schedule which shall henceforth be described in more detail.

\textit{Schedules} -- (s1) Our investigation started out with the double linear 
schedule of conventional PSSA, exactly as described in the last paragraph of 
Sect.~\ref{sec:PSSA}.
(s2) In addition, we tried a single linear schedule which omits the second 
annealing phase of the double linear schedule and immediately jumps towards the 
terminal search phase of improved PSSA after completing $\maxiter/2$ annealing 
steps.
(s3) The third temperature schedule we tried is a double exponential schedule. 
Its total runtime, gets exactly the same amount of annealing steps $\maxiter$ 
as the original double linear schedule.
It further initiates the first and the second annealing phase at exactly the 
same temperatures $T_{0}$ and $T_{\maxiter/2}$ as the original double linear 
schedule.
However, rather than decreasing the temperature linearly, the temperature is 
decreased exponentially.
To this end the temperature is updated every $1000$ annealing steps by 
multiplying the current temperature with a cooling factor $\beta<1$.
A larger value of $\beta$ corresponds to a slower cooling rate and tends to 
give better results.
However, one has to ensure that the system is sufficiently cooled at the end of 
each annealing phase in order not to jump out of the optimal configuration.
Here, we used the cooling rate $\beta=0.9999$.
(s4) The fourth and last schedule we tried is a single exponential schedule 
which is identical to the double exponential schedule in the first annealing 
phase.
Subsequently it skips the second annealing phase and directly jumps to the 
terminal search phase of improved PSSA after completing $\maxiter/2$ annealing 
steps.
Finally, we remark that the schedules $p_{s}(t)$ and $p_{a}(t)$ which coordinate 
swap and shift proposals as well as shift directions, remain exactly as 
described in the last paragraph of Sect.~\ref{sec:PSSA}.

\begin{figure}[tb]
  \centering{\includegraphics[width=\linewidth]{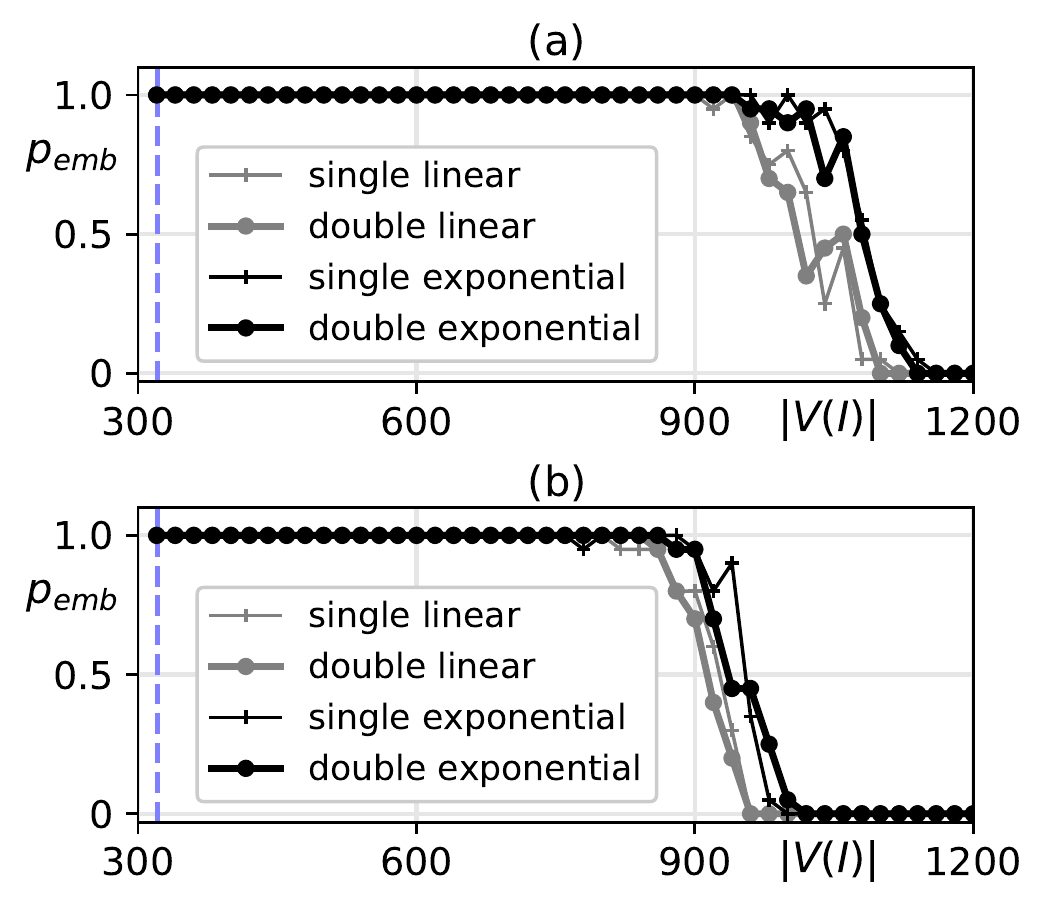}}
  \caption{Comparing the empirical embedding probability (20 inputs) of improved 
PSSA using single or double linear (thin/thick gray lines) as well as 
single or double exponential (thin/thick black lines) schedules for (a) 
random cubic and (b) Bar{\'a}basi-Albert type input graphs on a 
$\king_{320,320}$ hardware. A (dashed) vertical line indicates the embedding 
threshold $321$ of the best known complete graph embedding. Improved PSSA (a) 
includes (b) excludes degree-weighted shifts.}
	\label{fig:schedules}
\end{figure}
Admittedly, our methodology of improving the embedding performance based on the 
functional form of the annealing schedule is rather phenomenological.
Our motivation for doing so originates from our experience with simulated 
annealing on large spin systems, where we observed that exponential schedules 
tend to perform better than linear ones.
In a similar manner our methodology is justified by the phenomenological results 
depicted in Fig.~\ref{fig:schedules}.
In this figure we compare the performance of improved PSSA using the four 
different temperature schedules (s1-s4) for embedding (a) random cubic graphs 
and (b) Bar{\'a}basi-Albert random graphs into a King's graph $\king_{320,320}$.
Our results corroborate that the temperature schedules from the exponential 
family (black lines) tend to outperform the linear temperature schedules 
(gray lines), both on (a) random cubic  and (b) Bar{\'a}basi-Albert random 
graphs.
In addition we observe that the performance of single and double linear as well 
as single and double exponential schedules, respectively, is identical within 
the bounds of statistical fluctuations.
Meanwhile it has remained impossible to find a single instance of a random 
Erd\H{o}s-R{\'e}nyi graph with 20\% edge density and more than 321 vertices 
which could be embedded into the King's graphs $\king_{320,320}$ even with 
exponential scheduling.

\section{Embedding performance on increasing hardware}
\label{sec:Scalability}

Finally, we evaluate the empirical embedding threshold $\thres(L)$ of PSSA on 
hardware graphs of increasing size.
To this end we proceed as follows:\
(i) We fix the size of the hardware by setting the parameter $L=20$.
(ii) Starting from $|\vertices(\input)|=L$ we successively increase the size of 
the input problem.
For each pair $(L,|\vertices(\input)|)$ we create 20 input samples $\input_{s}, 
s=1,\cdots,20$ and try to embed them into $\king_{L,L}$ using (improved) PSSA.
If at least 19 out of 20 input graphs could successfully be embedded we proceed 
by increasing the number vertices $|\vertices(\input)|$.
The first time we find a pair $(L,|\vertices(\input)|)$ for which less than 19 
out of 20 cases were embeddable, we record $|\vertices(\input)|$ as the 
embedding threshold $\thres(L)$ and reinitiate step (ii) on a larger King's 
graph $L\gets L+20$.
We stop, if $L>320$.

\begin{figure}[tb]
  \centering{\includegraphics[width=\linewidth]{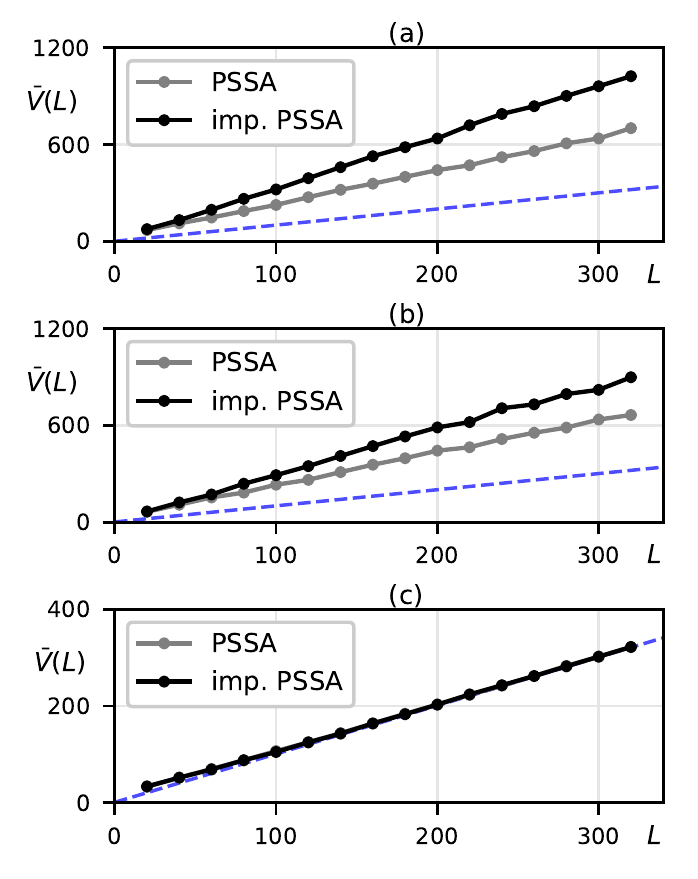}}
  \caption{Embedding threshold $\thres(L)$ of PSSA (gray) and improved PSSA 
(black) for increasing hardware King's graph $\king_{L,L}$ for (a) random cubic, 
(b) Bar{\'a}basi-Albert, and (c) Erd\H{o}s-R{\'e}nyi type input graphs. A dashed 
(blue) line indicates the minimal embedding threshold ensured by the existence 
of the best known complete graph embedding.}
  \label{fig:scaling}
\end{figure}
The results are shown in Fig.~\ref{fig:scaling} which depicts the embedding 
threshold $\thres(L)$ up to which (improved) PSSA finds minor embeddings with 
95\% embedding probability on hardware King's graph $\king_{L,L}$ of increasing 
size $L$.
For (a) random cubic and (b) Bar{\'a}basi-Albert type input graphs we 
phenomenologically observe a linear scaling of $\thres(L)$ with $L$.
In both cases $\thres(L)$ approximately scales as $\thres(L)= c \times L$ 
with (a) $c=3.2$ and (b) $c=2.8$.
Thus it maintains a clear advantage over the deterministic minor embedding of 
$\complete_{N}$ for which $\thres(L)=L+1$.
In addition, this constitutes an average performance gain of 42\% and 28\%, 
respectively, over the previously published version of PSSA \citep{Sug2018}.
On the other hand, for (c) Erd\H{o}s-R{\'e}nyi type input graphs with 20\% edge 
density $\thres(L)$ is hardly larger than $L+1$ even for small hardware graphs 
$\king_{L,L}$, eventually approaching the threshold of the best known complete 
graph embedding $\complete_{L+1}$ as $L$ gets larger.
The results in Fig.~\ref{fig:scaling} use improved PSSA with terminal search 
phase and double exponential schedules.
Degree-weighted shifts are applied only for random cubic graphs.
A discussion of these results will be given in Section~\ref{sec:Summary}.

\section{Can the best known complete graph embedding be improved?}
\label{sec:CliqueMinorScaling}

The results of the previous section show that PSSA can maintain an embedding 
threshold $\thres(L)$ which exceeds the minimal embedding performance ensured by 
the best known complete graph embedding for certain inputs even on large 
hardware graphs.
Yet, it also shows that the embedding threshold of Erd\H{o}s-R{\'e}nyi graphs 
with constant edge density cannot exceed the minimal embedding performance 
ensured by the best known complete graph embedding on large hardware graphs.
This result emphasizes the outstanding role of the best known complete graph 
embedding as a baseline embedding which ensures a minimal embedding performance.
This naturally raises two questions.
(i) Can a King's graph host minor embeddings of larger complete graphs?
(ii) Can the hardware graph be optimized to host larger complete graphs?

In this section we address both issues in part.
(i) We use the concept of treewidth to show that a King's graph $\king_{L,L}$ 
cannot contain the minor embeddings of a complete graph $\complete_{N}$ with 
$N>2L$ vertices.
We believe this bound is still quite loose but cannot prove that the best known 
complete graph embedding hosting input graphs $\complete_{N}$ with $N=L+1$ 
vertices is (close to) optimal.
(ii) We show that any hardware graph $\hardware$ with bounded coordination 
number $d$ (a common restriction for quantum and CMOS annealers) can, at most, 
embed complete graphs with $\mathcal{O}(\sqrt{|\vertices(\hardware)|})$ 
vertices.
Both the King's graph and the Chimera graph take this order and are thus in 
some sense optimal.
Note that the style of the paper will henceforth become more mathematical.
A reader who is more interested in the discussion of our improved PSSA is 
advised to skip to the summary section.

\subsection{Upper bound on complete graphs embeddable into a King's graph}

In this section we prove that a complete graph $\complete_N$ with $N$ vertices 
cannot be embedded into a King's graph $\king_{L,L}$, if $N>2L$.
The cornerstone of our proof is the following property, based on the concept of 
treewidth, which will be defined further below.
\begin{property}[\cite{Halin1976}]
\label{prop:MinorInclusion}
Let $\input$ and $\hardware$ be two graphs. 
If $\input$ is a minor of $\hardware$, its treewidth $\tw(\input)$ is smaller 
or equal than the treewidth $\tw(\hardware)$ of $\hardware$
\begin{equation}
  \label{eq:minortw}
 \input \text{ being a minor of } \hardware \Rightarrow 
\tw(\input)\le\tw(\hardware).
\end{equation}
\end{property}
We then exploit this property by identifying the input with a complete graph 
$\input = \complete_N$, whose treewidth is known $\tw(\complete_N)=N-1$ 
\citep{Fom2010}.
Finally, we identify $\hardware$ with a King's graph $\king_{L,L}$ and prove the 
following upper bound on the treewidth of a King's graph further below
\begin{equation}
 \label{eq:twking}
 \tw(\king_{L,L}) \le 2L-1.
\end{equation}
Incidentally, this implies that a complete graph $\complete_N$ with vertices 
$N>2L$ cannot be minor embedded into a King's graph $\king_{L,L}$, by means of 
property~\ref{prop:MinorInclusion}.

In order to prove Eq.~\eqref{eq:twking} we require some additional definitions.
Let $\graph$ be a general graph.
A path of the graph $\graph$ is a sequence of vertices $\langle v_1, v_2, v_3, 
\cdots, v_n \rangle$ without any repetition such that $(v_l, v_{l+1}) \in 
\edges(\graph)$ holds for all $l=1, \cdots, n-1$.
A cycle of the graph $\graph$ is a sequence of vertices without any repetition 
$\langle v_1, v_2, v_3, \cdots, v_n \rangle$ such that $( v_n, v_1 ) \in 
\edges(\graph)$ and $( v_l, v_{l+1} ) \in \edges(\graph)$ holds for all $l=1, 
\cdots, n-1$.
A graph $\graph$ is connected, if for all the pairs of vertices $v, v' \; ( v 
\neq v' )$, there is a path connecting $v$ and $v'$.
A tree $\tree$ is a connected graph without any cycle.
Note that if $\graph$ is a tree, there exists a unique path for each pair of 
vertices in $\vertices(\graph)$.

\begin{definition}[Tree Decomposition \citep{Robertson1986}]
A tree decomposition of $\graph$ is a family $\{ X_i \}_{i \in \ind}$ of 
vertex subsets $X_i\subseteq\vertices(\graph)$ together with a tree graph 
$\tree$ connecting the indices $i\in \ind$ of the subsets $X_{i}$, such that 
the following properties hold.
The vertex subsets $X_i\subseteq\vertices(\graph)$ cover the vertices and edges 
of 
the graph $\graph$ according to
\begin{enumerate}
  \item[] \hspace{-0.5cm}\treedec{1} $\bigcup_{i \in \ind} X_i = \vertices ( 
\graph)$,
  \item[] \hspace{-0.5cm}\treedec{2} for every edge $( v, v' ) \in \edges 
(\graph)$, there exists $i \in \ind$ such that $v, v' \in X_i$ holds.
\end{enumerate}
In addition we require that
\begin{enumerate}
  \item[] \hspace{-0.5cm}\treedec{3} for all $i, j, k \in \ind$, if $j$ is on 
the 
path of $\tree$ from $i$ to $k$, then $X_i \cap X_k \subseteq X_j$. 
\label{condition3}
\end{enumerate}
\end{definition}
Subsequently, for each tree decomposition, we define its width as $\max_{i \in 
\ind} | X_i | - 1$ through the subset $X_i$ which contains the maximal amount 
of 
vertices $| X_i |$.
Finally, the treewidth of a graph $\graph$ is the minimal width among all the 
possible tree decompositions of $\graph$.
The treewidth is denoted by $\tw (\graph)$.

\begin{figure}[tb]
  \centering{\includegraphics[width=0.8\linewidth]{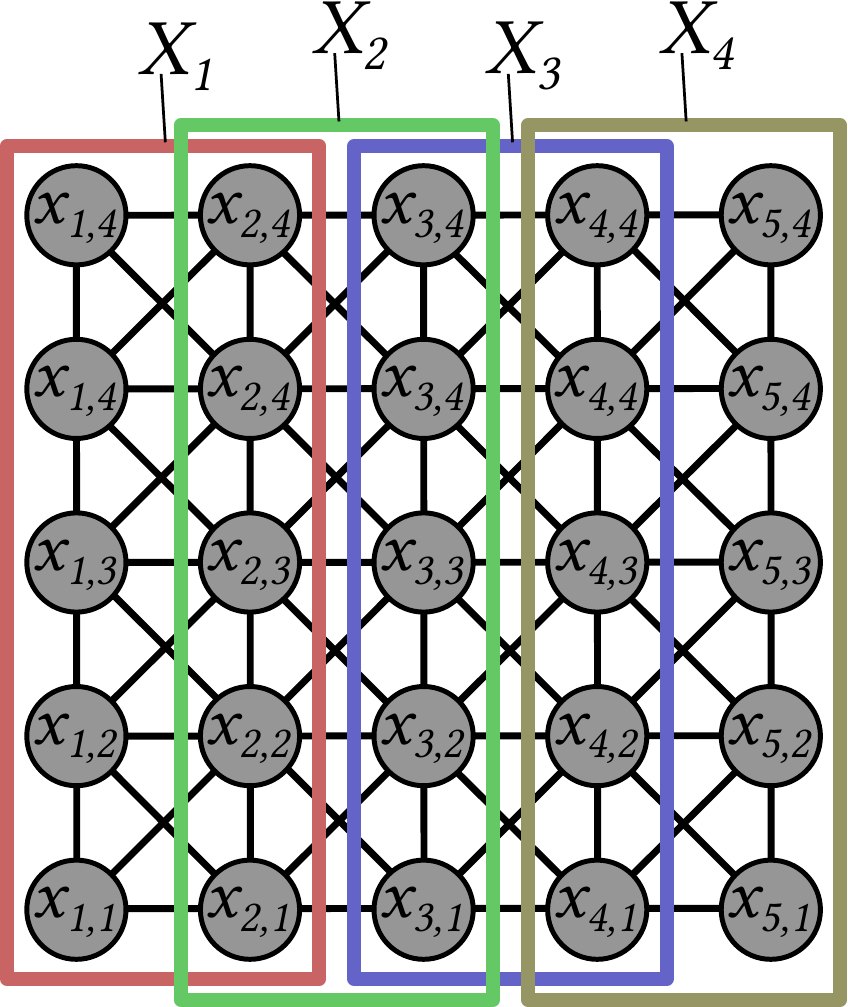}}
  \caption{Illustration of the proposed tree decomposition for constructing the 
upper bound on the treewidth of a King's graph $\king_{L,L}$ using the example 
$L=5$. The corresponding sets $X_{i}$ for $i=1,2,3,4$ are indicated by boxes 
with thick fat boundaries.}
  \label{fig:treewidth}
\end{figure}

\textit{Upper bound for treewidth of King's graph} -- We now prove 
Eq.~\eqref{eq:twking} by constructing a concrete tree decomposition of a King's 
graph and computing its width.
To this end, let $\king_{L,L}$ be the King's graph of size $L \times L$ and 
denote its vertex set as
\begin{align}
\vertices(\king_{L,L}) = \{x_{i,j}\quad|\quad i,j=1,\cdots,L\},
\end{align}
where $x_{i,j}$ is the vertex located at the point $( i, j )$ in the plane.
See Fig.~\ref{fig:treewidth} for an example.

We construct a tree decomposition of the King's graph $\king_{L,L}$ as follows:\
Let $\{ X_i \}_{i \in \ind}$ be a family of vertex subsets
\begin{equation}
X_i = \left\{ x_{i,j}, x_{i+1,j} \,|\, j=1,\cdots,L \right\},
\end{equation}
with  $i\in \ind = \{ 1, \cdots, L-1 \}$.
Let $\tree$ be a tree on the index set ($\vertices(\tree)=\ind$) with edges 
$\edges(\tree) = \{ ( 1, 2 ), ( 2, 3 ), \cdots, ( L-2, L-1 ) \}$ forming a 
simple path graph $\langle1,2,3,...,L-1\rangle$ on the indices $i$ of $X_{i}$.
See Fig.~\ref{fig:treewidth} for an example.
Then, the family of the subsets $\{ X_i \}_{i \in \ind}$ along with the tree 
$\tree$ is a tree decomposition of $\king_{L,L}$.
This can be checked as follows:
\treedec{1} holds because $\bigcup_{i \in \ind} X_i = \vertices(\king_{L,L})$.
\treedec{2} holds because (i) all horizontal edges $(x_{i,j}, x_{i+1,j})$ with 
$i=1,\cdots,L-1$ and $j=1,\cdots,L$ are contained in $X_i$, (ii) all diagonal 
$(x_{i,j}, x_{i+1,j+1})$ and anti-diagonal edges $(x_{i+1,j}, x_{i,j+1})$ with 
$i,j=1,\cdots,L-1$ are contained in $X_i$, (iii) vertical edges $(x_{i,j}, 
x_{i,j+1})$ with $i,j=1,\cdots,L-1$ are contained in $X_i$ while (iv) vertical 
edges $(x_{L,j}, x_{L,j+1})$ with $j=1,\cdots,L-1$ are contained in $X_{L-1}$.
\treedec{3} holds as well which can be seen as follows:\
(i) For indices $i$, $k$, with $|i-k|>1$ the intersection $X_i \cap X_k$ is 
empty and \treedec{3} is fulfilled.
(ii) For indices $|i-k|=1$ assume without loss of generality $i<k=i+1$. In 
this case $j$ is either $i$ or $k$ and the intersection $X_i \cap X_k = 
\left\{x_{i+1,j} \,|\, j=1,\cdots,L \right\}$ which is both a subset of $X_{i}$ 
and $X_{k=i+1}$.
(iii) Finally, consider the case $|i-k|=0$. In this case $j=i=k$ and $X_i \cap 
X_k \subseteq X_j$ is also true.
Altogether this shows that our above definition is a proper tree decomposition.

To complete the proof of Eq.~\eqref{eq:twking} we compute the width of the tree 
decomposition, given as $\max_{i \in \ind} | X_i | - 1 = 2 L - 1$, since all 
sets 
in the decomposition contain $2L$ vertices, i.e. $|X_i|=2L$ $\forall i\in 
\ind$.
By the definition of the treewidth, we conclude that $\tw(\king_{L,L} ) \le 2 L 
- 1 $. 

\subsection{Embedding complete graphs into hardware with fixed coordination 
number}
\label{sec:ResourceScaling}

We now touch upon the question of constructing an alternative hardware graph, 
which may potentially embed larger complete graphs.

\textit{Hardware graph with a fixed coordination number} -- To start with, let 
$\hardware$ be a connected hardware graph.
Since we are evaluating the question of an optimal hardware graph for the 
embedding of complete graphs, we will not yet fix its concrete structure.
For the moment, we only specify a fixed upper bound on the coordination number 
$d$ for each vertex of the hardware graph $\hardware$.
We believe that a fixed upper bound $d$ on the coordination number of each 
vertex is a reasonable hardware restriction which cannot easily be overcome in 
the foreseeable future.
In particular, current quantum annealers \citep{DWave} can hardly couple more 
than a few super-conducting quantum bits due to technical restrictions.
On the other hand, CMOS annealers \citep{Oku2017, Tak2019} exploit the 
sparsity of the hardware graph for parallel simultaneous spin updates and thus 
require a fixed upper bound on the coordination number.
In order to avoid pathological cases we assume $d>2$.
(If $d=2$, $\hardware$ could at most be a cycle or a path graph).

We now show that hardware graphs with fixed upper bound $d$ on the coordination 
number require a minimum of 
\begin{equation}
  \label{eq:MinimalVN}
 |\vertices(\hardware)| \ge \frac{N(N-3)}{d-2}
\end{equation}
vertices to host a minor of a complete graph $\complete_{N}$ with $N$ vertices.
In order to prove this, recall that a complete graph $\complete_{N}$ requires 
connecting each of its vertices to all of its other $N-1$ vertices.
However, a single vertex of the hardware graph can connect to at most $d$ 
neighbors and can thus in general not be connected to $N-1$ neighbors.
On the other hand, a super vertex comprising a total of $S$ vertices on the 
hardware can be connected to many more hardware vertices.
More specifically, a super vertex with $S$ vertices has $S\times d$ incoming 
edges.
In order to keep the super vertex connected we have to connect its vertices 
with at least $(S-1)$ internal edges, which requires sacrificaing at least 
$2(S-1)$ incoming edges of the super vertex.
(Note that each internal edge is incident to two vertices and thus equates to 
two incoming edges.)
Finally, utilizing the remaining incoming edges (at most $S(d-2)+2$) to connect 
the super vertex to at least $N-1$ hardware vertices which represent all other 
vertices of the original complete graph $\complete_N$ requires $S(d-2) + 2 \ge 
(N-1)$.
This results in a minimal size of a super vertex according to
\begin{equation}
  \label{eq:MinimalSupervertex}
  S\ge\frac{N-3}{d-2}.
\end{equation}
Further embedding a complete graph $\complete_N$ requires the super vertices of 
all $N$ vertices to take the aforementioned minimal size, thus, resulting in a 
hardware requirement as specified in Eq.~\eqref{eq:MinimalVN}.

Incidentally, the forgoing discussion demonstrates that for hardware graphs with 
fixed upper bound on the coordination number, the hardware spin resources 
$|\vertices(\hardware)|$ are of order $\Omega(N^2)$.
On the other hand, the number of vertices $N$ of the largest embeddable complete 
graph is inevitably of order $\mathcal{O}(\sqrt{|\vertices(\hardware)|})$.
Both for King's graphs and for Chimera graphs \citep{KlySulHum2014} the best 
known complete graph embeddings fulfill the above orders and are thus optimal in 
a certain sense.
On the other hand, it remains an interesting question for future research, if 
graph structures turning Eq.~\ref{eq:MinimalVN} into an equality exists for 
arbitrary $d$.
If so, it would pave the way for constructing hardware graphs with more 
efficient resource utilization with respect to complete graphs.

\section{Summary}
\label{sec:Summary}

We presented an improved version of PSSA and used it to evaluate the performance 
of embedding heuristics on hardware King's graphs of unprecedented size of 
102,400 spins as released with the latest CMOS annealing processor 
\citep{AC2018}.
The algorithmic improvements of PSSA included (i) an additional search phase, 
(ii) a degree-oriented super-vertex shift rule, and (iii) optimized annealing 
schedules.
The embedding performance of the improved PSSA was investigated with respect to 
various types of input graphs for hardware graphs of increasing size.
An embedding performance consistently exceeding the embedding threshold of the 
best known complete graph embedding by a factor of $c>1$ was observed for 
random-cubic ($c=3.2$) and Bar{\'a}basi-Albert ($c=2.8$) graphs.
This constitutes an average performance gain of 42\% and 28\%, respectively, 
over the previously published version of PSSA \citep{Sug2018}.
On the other hand, for sparse random graphs with constant edge density we 
observe that even the improved PSSA cannot exceed the embedding threshold of 
the best known complete graph embedding on large input graphs.
Finally, we derived a new upper bound on the vertex number of complete graphs 
embeddable into a hardware King's graph and showed that its size attains the 
maximal attainable order on hardware structures with a fixed coordination 
number.

\subsection{Discussion}

We now discuss several open questions of our research.
(i) Our results demonstrate that the improved PSSA can outperform the best known 
complete graph embedding on large hardware graphs for certain inputs, such as 
random cubic and Bar{\'a}basi-Albert graphs.
Yet, it remains a future task to evaluate whether the improved PSSA would also 
outperform the best known complete graph embedding for input graphs induced by 
concrete applications in quadratically unconstrained binary optimization.
We believe that graph coloring problems or solving the clique problem 
\citep{Luc2014} on social networks are suitable candidate applications.
(ii) It is currently unclear whether the improved PSSA performs close or far 
from optimality.
To address this question it would be desirable to compare its embedding 
threshold to the optimal embedding threshold attainable by means of exact 
algorithms \citep{Adl2010},
\begin{equation}
 \thres(\hardware, \graphclass, \embeddingalgorithm_{iPSSA}, p) \le 
 \thres(\hardware, \graphclass, \embeddingalgorithm_{exact}, p).
\end{equation}
This would allow for evaluating the headroom for further improvements of PSSA.
(iii) The results of Sect.~\ref{sec:CliqueMinorScaling} show that current 
hardware graphs are close to optimal for representing complete graphs.
Yet, it remains a question of future research, if hardware graphs $\hardware$ 
could be optimized to increase the embedding threshold $\thres(\hardware, 
\graphclass, \embeddingalgorithm, p)$ for certain classes $\graphclass$ of input 
graphs.
(iv) It remains a task for future research to develop easily accessible criteria 
which indicate whether an embedding heuristic such as improved PSSA can 
outperform the best known complete graph embedding for increasing hardware 
graphs on a certain class of input problems $\graphclass$.
Empirically, we observed that PSSA can outperform the best known complete graph 
embedding on large hardware graphs for random cubic and Bar{\'a}basi-Albert 
graphs whose edge set grows only linearly in the number of vertices 
$|\edges(\input)| \propto| \vertices(\input)|$.
On the other hand, we observed that  PSSA cannot outperform the best known 
complete graph embedding on large hardware graphs for Erd\H{o}s-R{\'e}nyi graphs 
whose thicker edge set grows quadratically with the number of vertices 
$|\edges(\input)| \propto |\vertices(\input)|^2$.
Yet, it is by no means clear, that the reverse should hold.
That is, we cannot prove that the number of edges scaling linearly in the size 
of the vertex set would guarantee embedding performance superior to the best 
known complete graph embedding.
Similarly, it is unclear, if the number of edges growing quadratically in the 
number of vertices necessarily implies that the embedding threshold shrinks to 
the performance of the best known complete graph embedding as the hardware 
increases.
(v) The main objective of improved PSSA is a high embedding threshold which 
outperforms the best known complete graph embedding even on large hardware 
graphs.
In particular for input problems where this objective fails, it is advisable 
to optimize the minor embedding heuristic with respect to other objectives.
Such an objective could, for example, be the runtime of the embedding heuristic 
\citep{Goo2018}.
In addition, a good minor heuristic should attempt to minimize the size of its 
super vertices to avoid invalid solutions in the subsequent annealing phase 
\citep{Ven2014, Boo2016}.
This secondary objective has partially been addressed by the cleanup process in 
the terminal search phase of our improved PSSA.
Investigating its effect in terms of precise data on the final distribution of 
super vertex sizes, before and after the cleanup of improved PSSA, remains a 
subject of future research.

\begin{acknowledgements}
We thank Hirofumi Suzuki, Kazuhiro Kurita, and Shoya Takahashi for supporting 
the organization of the ``Hokkaido University \& Hitachi 2nd New-Concept 
Computing Contest 2017'' and Ko Fujizawa and Masumitsu Aoki for stimulating 
discussions.
\end{acknowledgements}

\bibliographystyle{spbasic}      
\bibliography{bibliography}   

\section{Appendix}

\subsection{Experimental conditions}
\label{app:experimentalconditions}

\textit{Random cubic} graphs are generated using Algorithm 1 from 
\citep{Ste1999} for a target of $|\vertices(\input)|$ vertices of degree $d=3$.
Note that this implies that the number of edges grow as $|\edges(\input)| = 
3|\vertices(\input)|/2$.

\textit{Bar{\'a}basi-Albert} graphs are generated as described in 
\citep{Alb2002}.
In order to keep the graph connected we start out with $m_0(=2)$ connected 
vertices.
We then succesively add new vertices to the graph.
Every time we add a new vertex we connect it with the existing vertices by 
adding $m (= 2\le m_0)$ edges based on preferential attachement.
The proceedure is repeated until the graph has reached a prescribed number of 
vertices $|\vertices(\input)|$.
Note that the final graph has $|\edges(\input)| = m_0(m_0-1)/2 + m 
(|\vertices(\input)| - m_0)$ edges.

\textit{Erd\H{o}s-R{\'e}nyi} graphs are created by growing a tree up to the 
desired number of vertices.
To this end we add one vertex at a time and connect it to one of the existing 
vertices with equal probability.
Subsequently, we add edges to the tree by filling unoccupied edges with equal 
probability until the prescribed edge density $\rho$ is reached.
The slight modification from the standard Erd\H{o}s-R{\'e}nyi procedure ensures 
that the resulting random graph is connected.
Note that the number of edges of our random graphs grows as $|\edges(\input)| = 
\rho |\vertices(\input)|(|\vertices(\input)|-1)/2$.
In this paper we always set $\rho=0.2$ using an edge density of 20\%.
Our results on the scalability do not crucially depend on the precise value of 
$\rho$, that is, for large hardware graphs PSSA cannot improve over the 
embedding performance of the best known complete graph embedding, also for other 
values of $\rho$.

\textit{Schedule parameters} -- All data in this paper is produced using a 
maximum of $\maxiter = 7\times10^{7}$ iterations and initial temperatures 
$T_{0}=60.315$ and $T_{\maxiter/2}=33.435$. Linear schedules always terminate 
each annealing phase with temperature $T=0$. Exponential schedules use a 
cooling 
coefficient $\beta=0.9999$ to update the temperature every 1000 annealing steps 
as $T\gets\beta*T$. Swap and shift proposals are scheduled with linear 
schedules 
using $p_{s}(0)=1, p_{s}(\maxiter)=0, p_{a}(0)=0.095, p_{a}(\maxiter)=0.487$.

\subsection{Implementation of the terminal search phase}
\label{app:Implementation}

In this appendix we describe possible implementations of the subroutines 
$\isdeletable(\cdot)$ and $\bfspath(\cdot)$ as required in the terminal search 
phase of the improved PSSA.

\begin{figure}[tb]
  \centering{\includegraphics[width=0.85\linewidth]{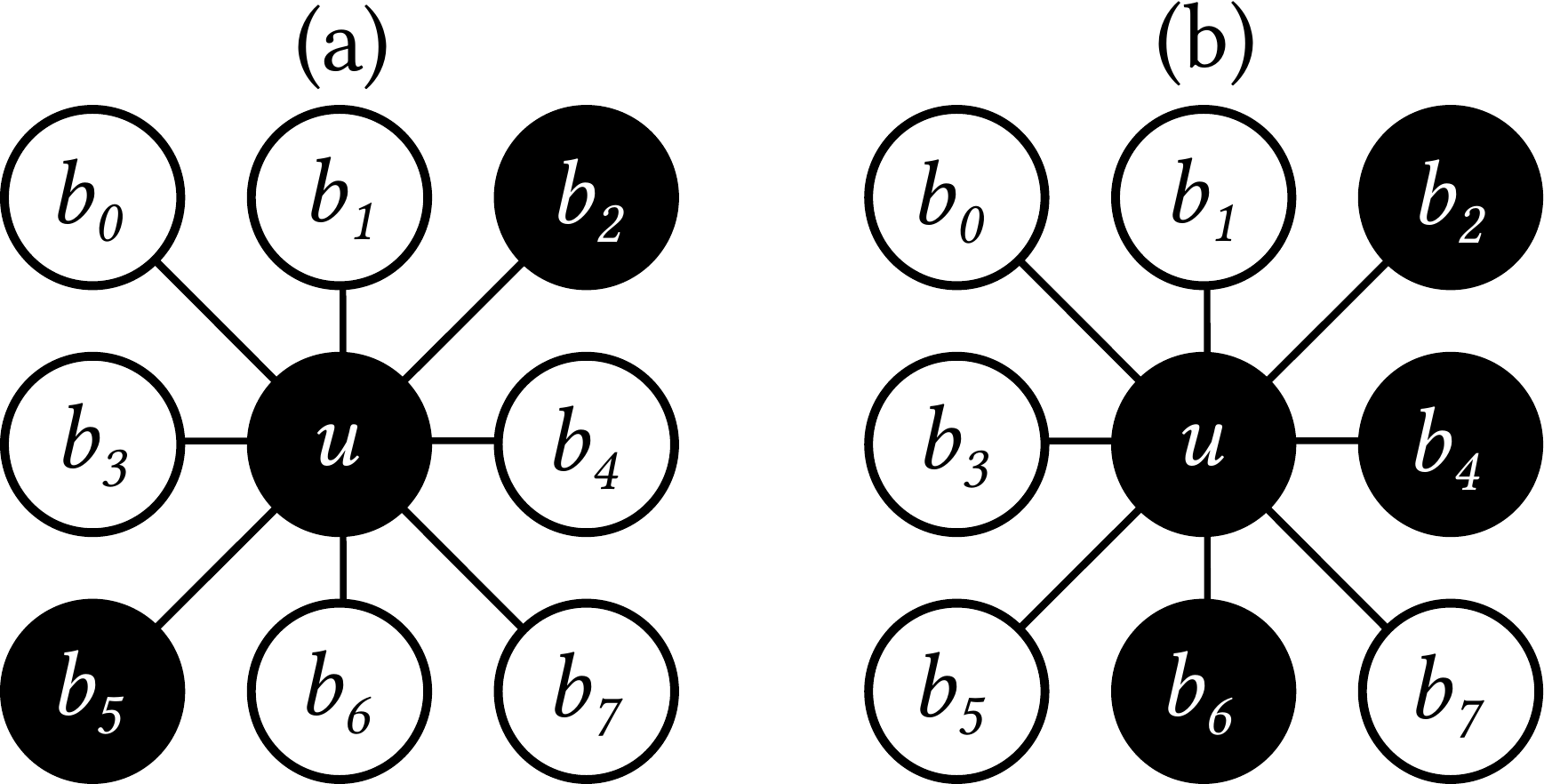}}
  \caption{Illustration of the bit pattern encoding the neighborhood of $u \in 
\vertices(\hardware)$. Bits $b_{j}$ of neighboring vertices belonging to the 
same super vertex as $u$ are set to $1$ and shown in black, while bits 
belonging 
to a super vertex different from $u$ are set to $0$ and shown in white. Bit 
patterns for which (a) $u$ is not deletable and (b) $u$ is deletable from its 
super vertex without destroying property \minor{1}, respectively. 
}
\label{fig:bitpattern}
\end{figure}

The subroutine $\isdeletable(\cdot)$ returns that a hardware vertex $u \in 
\vertices(\hardware)$ is deletable from its super vertex $\phi(i)$, if both
(p1) deleting $u$ from $\phi(i)$ does not destroy the non-empty connected 
structure of the super vertex \minor{1} and
(p2) deleting $u$ from $\phi(i)$ does not decrease the number of embedded edges 
$\embedded(\phi)$.
The subroutine $\isdeletable(\cdot)$ checks for property (p1) by encoding the 
super vertex occupation of the 8 neighboring vertices $v_{0}, \cdots, v_{7} \in 
\vertices(\hardware)$ of $u$ with $(u,v_{\nu})\in\edges(\hardware)$ into a 
pattern of 8 bits $b_{\nu}, \nu=0,\cdots,7$.
See Fig.~\ref{fig:bitpattern}.
In particular, the $\nu$th bit is set to $1$, if $v_{\nu}$ belongs to the same 
super vertex $\phi(i)$ as $u$ and $0$ otherwise
\begin{eqnarray}
b_{\nu} &= \begin{cases} 1 & \quad \text{if } v_{\nu} \in \phi(i) \text{ with } 
u \in \phi(i),\\
                         0 & \quad \text{if } v_{\nu} \notin \phi(i) \text{ 
with } u \in \phi(i). 
         \end{cases}
\end{eqnarray}
The subroutine $\isdeletable(\cdot)$ then checks property (p1) by computing the 
bit pattern of the current vertex $u$ and comparing it to a precomputed list of 
deletable patterns.
Deletable patterns have been obtained by inspecting all the 256 possible bit 
patterns and hard coding the deletable ones into the subroutine.
If the bit pattern of the current vertex $u$ is not in the list of deletable 
bit patterns, the subroutine decides that $u$ is not deletable and returns.
If the bit pattern of the current vertex $u$ is in the list of deletable bit 
patterns, the subroutine proceeds to checking (p2).

\begin{figure}[tb]
  \centering{\includegraphics[width=0.85\linewidth]{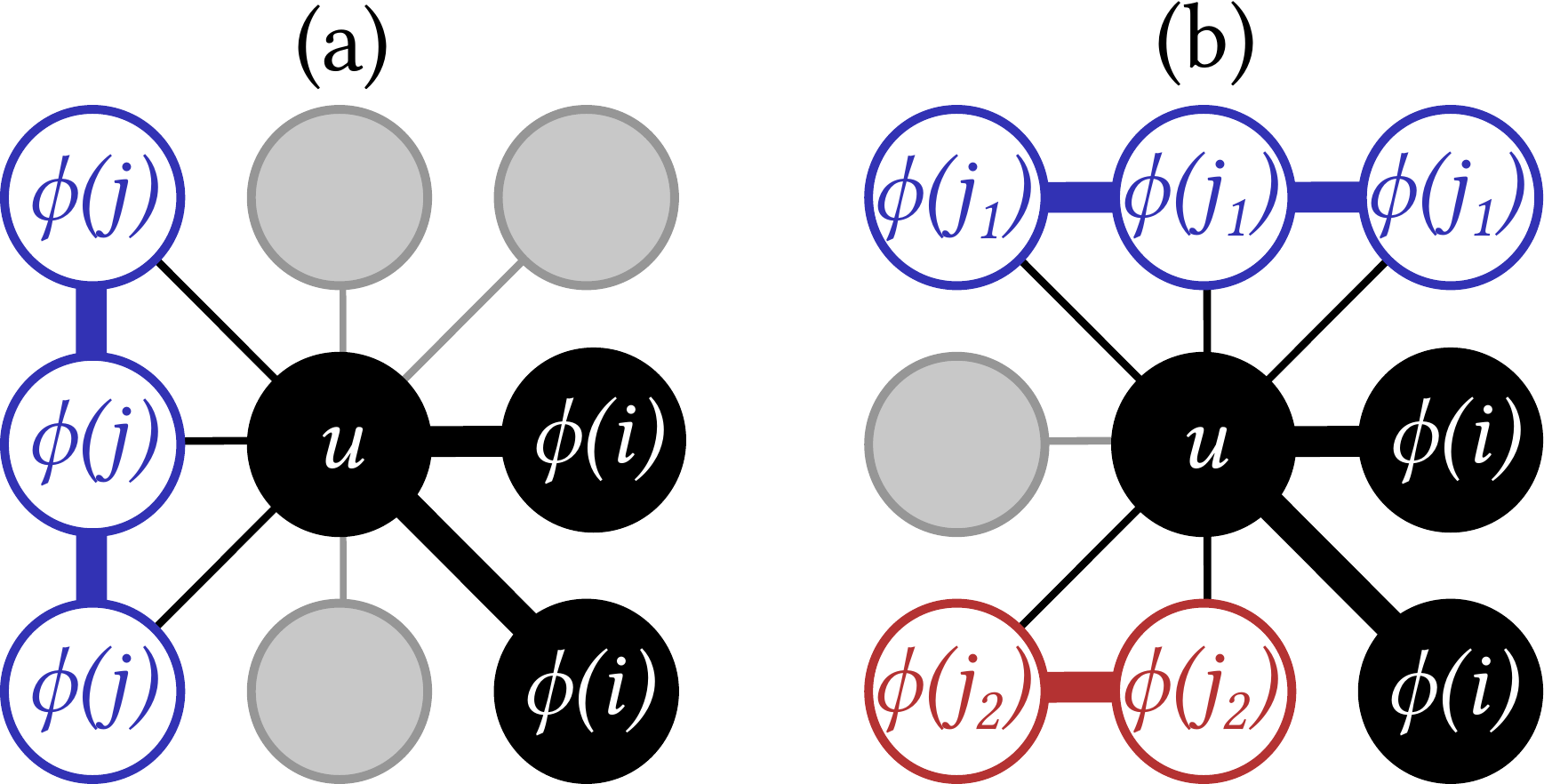}}
  \caption{Illustration of (a) a non-deletable and (b) a deletable vertex $u$ 
in 
the sense of (p2).
(a) Checking the neighborhood of the vertex $u\in\phi(i)$ gives $n(i,j)=3$ 
hardware representations of the edge $(i,j) \in \edges(\input)$ and five 
vertices and edges (gray and black), respectively, which do not represent an 
edge of the input graph.
Assuming that $\phi(i)$ and $\phi(j)$ have no further links elsewhere on the 
hardware, results in $u$ being not-deletable.
(b) Checking the neighborhood of the vertex $u\in\phi(i)$ gives $n(i,j_{1})=3$ 
hardware representations of the edge $(i,j_1)\in\edges(\input)$ and 
$n(i,j_{2})=2$ hardware representations of the edge $(i,j_2)\in\edges(\input)$.
Furthermore, there are three vertices and edges, respectively, (gray and black)
which do not represent an edge of the input graph.
Since, $\phi(i)$ is linked with $\phi(j_1)$ by more than $n(i,j_{1})=3$ edges 
and $\phi(i)$ is linked with $\phi(j_2)$ by more than $n(i,j_{2})=2$ edges, $u$ 
is deletable from $\phi(i)$.
  }
  \label{fig:p2}
\end{figure}

To check if $u$ is deletable in the sense of (p2), the subroutine proceeds as 
follows.
First, it scans the 8 neighbors $v_{\nu}, \nu=0,\cdots,7$ of $u\in\phi(i)$ and 
checks the corresponding edges $(u, v_{\nu}) \in \edges(\hardware)$.
During this process the corresponding pre-images $j_{\nu} = \phi^{-1}(v_{\nu}) 
\in \vertices(\input), \nu=0,\cdots,7$ are computed.
If none of the pairs $(i, j_{\nu})$ represents an edge of the input graph 
$(i, j_{\nu})\in\edges(\input)$ the subroutine decides that $u$ is deletable 
and returns.
On the other hand, for every pair $(i, j_{\nu})\in\edges(\input)$ which 
represents an edge of the input graph the subroutine records the vertex 
$j_{\nu}\in\vertices(\input)$ in a unique and sorted list.
The number of occurrences $n(i, j_{\nu})$ is tracked in a corresponding list.
Finally, the subroutine checks if $\forall j_{\nu}$ in the unique and sorted 
list of vertices the amount of hardware representations $n(i,j_{\nu})$ of the 
edge $(i,j_{\nu})$ is smaller than the total number of hardware representations 
$N(i,j_{\nu})$.
If this condition is fulfilled, the subroutine updates $N(i,j_{\nu})\gets 
N(i,j_{\nu}) - n(i, j_{\nu})$ for all $\forall j_{\nu}$ in the unique and 
sorted list of vertices and returns with $u$ being deletable.
Otherwise, it returns with $u$ not being deletable.
Note that the data on the amount of hardware edges $N(i, j)$ representing the 
edge $(i, j) \in \edges(\input)$ on the hardware has to be computed before 
the subroutine $\isdeletable(\cdot)$ is called for the first time.
A sketch illustrating the check of deletability in the sense of (p2) is 
illustrated in Fig.~\ref{fig:p2}.

The subroutine $\bfspath(\cdot)$ will operate on the hardware graph 
$\hardware$.
It will try to create a hardware link between $\phi(i)$ and $\phi(j)$ using the 
breath first search algorithm described in \citep[pp 594-602]{Cor2009} with 
slight modifications, as follows:\
(i) The queue will be initialized by including all vertices of $\phi(i)$ into 
the queue, marking their current status as gray, i.e., in the queue.
The free hardware vertices $\unassigned$, Eq.~\eqref{eq:unassigned}, are 
initialized as white, i.e., unvisited.
The vertices of $\phi(j)$ are marked as green, i.e., target reached.
Finally, all other vertices are marked as red, denoting occupied vertices of 
the 
hardware.
(ii) The subroutine $\bfspath(\cdot)$ will then successively dequeue vertices 
$u$ from the queue and check all its adjacent nodes $v$ in $\hardware$.
If $v$ is an occupied or a visited node, it will be skipped.
If $v$ is an unvisited node, $v$ is enqueued, $u$ is registered as its parent 
and $v$'s status is set to gray, i.e., in the queue.
Finally, if $v$ is a target node with $v\in\phi(j)$, breadth first search is 
stopped and the subroutine proceeds to the cleanup.
After checking all nodes $v$ adjacent to $u$ in $\hardware$ the color of $u$ is 
updated to black, i.e., visited and breadth first search proceeds to dequeuing 
the next vertex from the queue.
(iii) If a target node $v\in\phi(j)$ is found, its parents are traced back to 
$\phi(i)$ and the corresponding path of vertices is removed from the the set of 
free hardware vertices $\unassigned$, Eq.~\eqref{eq:unassigned}, and assigned to 
the super vertex $\phi(i)$.
After doing so $\bfspath(\cdot)$ returns successfully and proceeds to the next 
edge $(i,j) \in \edges(\input)$ whose super vertices $\phi(i), \phi(j)$ are not 
yet linked on the hardware $\hardware$.
Alternatively, $\bfspath(\cdot)$ might return unsuccessfully with an empty 
queue without hitting a single target vertex.
In this case $\unassigned$ and $\phi$ remain unchanged.

\end{document}